\documentclass[12pt]{article}

\usepackage{a4,a4wide}

\usepackage{amsmath}
\usepackage{amsfonts}
\usepackage{amssymb}
\usepackage{fancybox}
\usepackage[dvips]{color}

\usepackage{color}
\definecolor{blue1}{rgb}{0.15,0.15,0.50}
\usepackage[
debug,dvipdfm,
colorlinks=true,
urlcolor=blue1,
anchorcolor=blue,
citecolor=cyan,
filecolor=blue,
linkcolor=blue1,
menucolor=blue,
pagecolor=blue,
linktocpage=true,
pageanchor=false,
hypertex
]{hyperref}    

\def\Tr{\mathrm{Tr}}

\def\half{{1\over2}}
\def\nn{\nonumber\\}

\def\={\stackrel{\bullet}{=}}

\def\({\left(}
\def\){\right)}

\def\Tr{\mathrm{Tr}}

\def\cN{{\cal N}}

\def\cP{{\cal P}}

\def\mbf{\mathbf}

\def\mf {\mathfrak }

\def \be {\begin{equation}}
\def \ee {\end{equation}}
\def \bea {\begin{eqnarray}}
\def \eea {\end{eqnarray}}
\def\beal#1{\begin{align}#1\end{align}}
\def \nn {\notag\\}

\begin{document}

\makeatletter
\renewcommand{\theequation}{%
\thesection.\arabic{equation}}
\@addtoreset{equation}{section}
\makeatother

\begin{titlepage}
\vspace{-3cm}
\title{
\begin{flushright}
\normalsize{ TIFR/TH/14-19\\
June 2014}
\end{flushright}
       \vspace{1.5cm}
More on BPS States in $\cN=4$ Supersymmetric Yang-Mills Theory on $\mathbf R \times \mathbf S^3$
       \vspace{1.5cm}
}
\author{
Shuichi Yokoyama
\\[25pt] 
\!\!\!\!\!\!\!\!{\it \normalsize Department of Theoretical Physics, Tata Institute of Fundamental Research,}\\
\!\!\!\!\!\!\!\!{\it \normalsize Homi Bhabha Road, Mumbai 400005, India}\\
\\[10pt]
\!\!\!\!\!\!\!\!{\small \tt E-mail: yokoyama(at)theory.tifr.res.in}
}

\date{}
\maketitle

\thispagestyle{empty}

\vspace{.2cm}

\begin{abstract}
\vspace{0.3cm}
\normalsize

We perform a systematic analysis on supersymmetric states in $\cN=4$ supersymmetric Yang-Mills theory (SYM) on 
$\mbf R\times \mbf S^3$. 
We find a new set of $1/16$ BPS equations and determine the precise configuration of the supersymmetric states 
by solving all $1/16$ BPS equations when they are valued in Cartan subalgebra of a gauge group and the fermionic fields vanish. 
We also determine the number of supersymmetries preserved by the supersymmetric states varying the parameters of the BPS solutions. 
As a byproduct we present the complete set of such supersymmetric states in $\cN=8$ SYM on $\mbf R\times \mbf S^2$ by carrying out dimensional reduction. 

\end{abstract}
\end{titlepage}

\section{Introduction} 
\label{introduction}

$\cN=4$ supersymmetric Yang-Mills theory (SYM) 
in four dimensions has been a fundamental theoretical tool to 
extract important lessons of duality. 
In particular $\cN=4$ SYM enjoys S-duality \cite{Osborn:1979tq,Sen:1994yi},  
which is a generalization of the electro-magnetic duality \cite{Goddard:1976qe,Montonen:1977sn},  
and also exhibits a dual description of type IIB supergravity on $\mbf{AdS}_5 \times \mbf S^5$ under a particular limit \cite{Maldacena:1997re}. 
For precise investigation of these kinds of strong-weak duality, 
supersymmetry often plays important roles. 

Since maximally supersymmetric completion of Yang-Mills theory 
gives rise to full quantum conformal invariance 
in the flat space \cite{Sohnius:1981sn,Brink:1982pd,Howe:1983sr,Seiberg:1988ur}, 
$\cN=4$ SYM can be studied by mapping the system onto $\mbf R\times \mbf S^3$ by a conformal transformation, 
which makes the system free from infra-red divergence and discretizes the spectrum of the system
with the mass gap of order the inverse radius of the three sphere. 
In particular discrete spectrum of the supersymmetric (or BPS) states of the theory has attracted 
a great deal of attention in the context of AdS$_5$/CFT$_4$ duality, 
which can be tested by confirming match of supersymmetric spectra in both sides \cite{Witten:1998qj,Gubser:1998bc}. 

One aspect in the study of BPS spectrum is counting BPS states. 
(Other works related to $\cN=4$ SYM on $\mbf R\times \mbf S^3$ are found, for instance, in \cite{Witten:1998zw,Okuyama:2002zn,Berenstein:2005aa,Berkooz:2008gc}.)
In particular an important tool to study BPS spectrum is a (superconformal) index \cite{Witten:1982df}, 
which encodes BPS spectrum as a form of polynomial. 
A superconformal index of the $\cN=4$ SYM has been computed exactly in arbitrary $SU(N)$ gauge group \cite{Kinney:2005ej}. (See also \cite{Romelsberger:2005eg,Imamura:2011uw,Nawata:2011un}.) 
Under the large $N$ limit with the charges kept finite 
the index of the $\cN=4$ SYM precisely matches 
that of multi-particle states of supergravity multiplet in type IIB supergravity 
on $\mbf{AdS}_5 \times \mbf S^5$ \cite{Kinney:2005ej}.
This result is reasonable in the sense that 
other supersymmetric objects such as (dual) BPS giant gravitons \cite{McGreevy:2000cw,Grisaru:2000zn,Hashimoto:2000zp} 
and BPS black holes \cite{Gutowski:2004yv,Gutowski:2004ez,Kunduri:2006ek} in the dual geometry have much larger charges, which are of order $N$, $N^2$ respectively. 
Counting of supersymmetric states with such large charges in the $\cN=4$ SYM
has also been done as a purely mathematical problem
by recasting BPS states as cohomology of a supercharge 
\cite{Grant:2008sk}.  
As a result, an index of $1/8$ BPS (dual) giant gravitons \cite{Mandal:2006tk,Biswas:2006tj} 
has been reproduced from the $\cN=4$ SYM,
whereas that of $1/16$ supersymmetric black holes has not. 
$1/16$ BPS states are less understood and expected to have a key to make more progress in AdS$_5$/CFT$_4$ duality.  
(See \cite{Chang:2013fba} for a recent study in this direction.) 

Under the circumference this paper is aimed at carrying out a systematic analysis on $1/16$ BPS states of the $\cN=4$ SYM.  
The goal of this paper is to clarify basic information on the BPS states of the $\cN=4$ SYM such as 
the $1/16$ BPS equations, the BPS configuration and preserved supersymmetries by setting assumptions that the fermionic fields vanish and that the fields are valued in Cartan subalgebra of a gauge group. 

The rest of this paper is organized in the following way. 
In \S \ref{N4SYM} we review basics of $\cN=4$ SYM explaining our convention. 
In \S \ref{BPSN=4SYM} we carry out a systematic analysis on supersymmetric states of the $\cN=4$ SYM
by setting the fermionic fields to zero.
In \S \ref{sphericalsym} we study the spherical symmetric BPS states for warm-up.
In \S \ref{nonsphericalBPS} we study supersymmetric states with angular momenta, which contain $1/16$ BPS states. 
In \S \ref{countSUSY} we count the number of preserved supersymmetries of the BPS states determined in \S \ref{nonsphericalBPS}. 
In \S \ref{BPSN=8SYM} we derive similar information of BPS states in $\cN=8$ SYM on $\mbf R\times \mbf S^2$ by performing dimensional reduction.  
\S \ref{discussion} is dedicated to discussion and future works. 
Appendix contains our convention (\ref{convension}), and basic information on $\mbf S^3$ (\ref{s3})
including construction of the scalar spherical harmonics on $\mbf S^3$ (\ref{harmonics})
and reduction of information of $\mbf S^3$ to that of $\mbf S^2$ (\ref{s2}). 

\section{$\cN=4$ SYM on $\mbf R\times \mbf S^3$}
\label{N4SYM}

In this section we review basics of $\cN=4$ SYM on $\mbf R\times \mbf S^3$ 
so as to be self-contained. 
The $\cN=4$ SYM consists of six real scalars, gauge field and 
their supersymmetric partners, 
all of which are valued in the adjoint representation of a gauge group.%
\footnote{In this paper we do not need to specify a gauge group.} 
This theory has $PSU(2,2|4)$ global symmetry, 
and especially $SO(6)_R \simeq SU(4)_R$ R-symmetry. 
The fermionic fields are in the anti-fundamental representation of $SU(4)_R$, 
which we denote by $\lambda_A$, 
and the six scalar fields form the anti-symmetric representation of $SU(4)_R$, 
which is denoted by $X^{AB}$ satisfying
\be
X^{AB} =-X^{BA}, \quad (X^{AB})^\dagger=\half \varepsilon_{ABCD}X^{CD}=:X_{AB},
\label{Xconstraint}
\ee
where $A, B=1,2,3,4$. 
The action of $\cN=4$ SYM on $\mbf R \times \mbf S^3$ is given by 
\beal{
S=& {1\over g^2} \int dt r^3d^3\Omega\; \Tr
\biggl[-{1\over 4} F_{\mu\nu} F^{\mu\nu}
-\half D_\mu X_{AB} D^\mu X^{AB}
+{1\over4}[X_{AB},X_{CD}][X^{AB},X^{CD}]
-{1\over2r^2} X_{AB} X^{AB} \nn
&\qquad\qquad+{i}(\lambda_A)^\dagger \gamma^\mu D_\mu \lambda_{A}
+ {\lambda_{A}} [X^{AB},  \lambda_{B}] +  {(\lambda_A)^\dagger} [X_{AB},  (\lambda_B)^\dagger]
\biggl] 
\label{N4SYMaction}
}
where $g$ is a gauge coupling, $r$ is the radius of the three sphere, $d^3\Omega={1\over 8} d\theta\sin\theta d\phi d\psi$ is the volume form of the unit three sphere, 
$D_\mu$ is the covariant derivative with respect to gauge and space index, 
which acts on the fields as follows.
\beal{
D_\mu X^{AB} =& \partial_\mu X^{AB} + i [A_\mu,  X^{AB}] \\
D_\mu \lambda_{A} 
=& \partial_\mu\lambda_A + {1\over4} \omega_{\mu,\nu\rho}\gamma^{\nu\rho}\lambda_A+ i [A_\mu,  \lambda_A]
} 
where $\omega_{\mu,\nu\rho}$ is the one form connection of the space-time.
See Appendix \ref{s3} for more details.

For later convenience 
we present equations of motion and conserved charges in the bosonic part. 
These can be computed in the standard way. 
Equations of motion for the gauge field and complex scalar fields in the bosonic part are  
\beal{
& D_\mu F^{\mu\nu} - i [X_{AB}, D^\nu X^{AB}]=0, \label{eomF}\\
& D^\mu D_\mu X^{AB} + [X_{CD},[X^{AB},X^{CD}]] - {1 \over r^2} X^{AB} =0. \label{eomX}
}
$SU(4)_R$ R-symmetry charge is 
\beal{
R^{C}{}_A =&i  \int d^3\Omega~   {1\over g^2} \Tr \( - X^{CB} D^0 X_{AB} + D^0 X^{CB} X_{AB} \). 
\label{Rcharge}
}
The energy is 
\beal{
H 
=&  \int d^3\Omega~  {1\over g^2} \Tr\(\half (F^{0i})^2+ {1\over 4} (F^{ij})^2 +\half |D_0 X_{AB}|^2 +\half |D_i X^{AB}|^2 + {1\over 4} |[X_{AB},X_{CD}]|^2 +{1\over 2r^2} |X_{AB}|^2 \) 
\label{energy}
}
where $|A|^2 = AA^\dagger$. 
The angular momentum is  
\beal{
J^{i} =&  \int d^3\Omega~   {1\over g^2} \Tr \(F^{0}{}_\mu F^{i\mu} +\half ( D^{0}X_{AB} D^{i} X^{AB}+D^{i}X_{AB} D^{0} X^{AB}) \) 
\label{momenta}
}
where $i = \theta, \phi, \psi$.

The action \eqref{N4SYMaction} is invariant under the following supersymmetry transformation rule. 
\beal{
&\Delta_\epsilon A_{\mu} 
= i ((\epsilon_A)^* \gamma_{\mu} \lambda_{A}
+{\epsilon_{A}} \gamma_{\mu} (\lambda_A)^\dagger), \nn
& 
\Delta_\epsilon X^{AB} = i (- \varepsilon^{ABCD} {\epsilon_C} \lambda_{D}
-  (\epsilon_A)^* (\lambda_B)^\dagger + (\epsilon_B)^* (\lambda_A)^\dagger ), 
\label{susy}\\
&\Delta_\epsilon \lambda_{A} 
= \half F_{{\mu}{\nu}}\gamma^{{\mu}{\nu}} \epsilon_{A}
- 2 D_\mu X_{AB} \gamma^\mu (\epsilon_B)^*
-2 i[X_{AB}, X^{BC}] \epsilon_{C}
-X_{AB} \gamma^\mu \nabla_\mu (\epsilon_B)^*, \notag
}
where $\nabla_\mu$ is the covariant derivative for spin
and $\epsilon_{A}$ is a supersymmetry parameter valued in Grassmann number,
which is given by 
a conformal Killing spinor on $\mbf R\times\mbf S^3$
\be
\nabla_{\mu}\epsilon^{(\pm)}_{A} = \pm {i \over 2r}\gamma_{\mu}\gamma^t \epsilon^{(\pm)}_{A}.
\label{KSequation}
\ee 
Let us solve this Killing spinor equation briefly. 
For $\epsilon^{(+)}_{A}$, \eqref{KSequation} reduces to 
\beal{
\partial_{t}\epsilon^{(+)}_A = {i \over 2r} \epsilon^{(+)}_A, \quad   \partial_{\psi}\epsilon^{(+)}_A = {i \over 2}\gamma_{3}{}^t  \epsilon^{(+)}_A, \quad
\partial_{\theta}\epsilon^{(+)}_A =0, \quad 
\partial_{\phi}\epsilon^{(+)}_A =0. \quad 
\label{KSeq+}
}
This can be solved as follows. 
\be
\epsilon^{(+)}_{A} = e^{{i \over 2r} t} e^{ - {i \over 2} \gamma_{30} \psi} \eta^{(+)}_{A}
\label{KS+}
\ee
where $\eta^{(+)}_{A}$ is a constant spinor. 
On the other hand, for $\epsilon^{(-)}_{A}$, \eqref{KSequation} becomes 
\be
\partial_{t}\epsilon^{(-)}_A = - {i \over 2r} \epsilon^{(-)}_A, \quad   
\nabla^{\mbf S^2}_{\theta}\epsilon^{(-)}_A = - {i \over r}\gamma_{{\theta}}{}^t \epsilon^{(-)}_A, \quad 
\nabla^{\mbf S^2}_{\phi}\epsilon^{(-)}_A = - {i \over r}\gamma_{{\phi}}{}^t \epsilon^{(-)}_A,\quad
\partial_{\psi}\epsilon^{(-)}_A =0,
\label{KSeq-}
\ee
where $\nabla^{\mbf S^2}_\mu$ is the spin covariant derivative on $\mbf S^2$.
More explicitly, the covariant derivatives act on a spinor as follows.  
\be
\nabla_\theta \epsilon^{(-)}_{A}= \partial_\theta \epsilon^{(-)}_{A}, \quad
\nabla_\phi \epsilon^{(-)}_{A}= \partial_\phi \epsilon^{(-)}_{A}+\half \cos\theta \gamma^{21}\epsilon^{(-)}_{A}.
\ee
The solution of Killing spinor equation is 
\be
\epsilon^{(-)}_{A} = e^{- {i \over 2r} t} e^{ - {i \over 2} \gamma_{01} \theta} e^{- \half \gamma_{21} \phi} \eta^{(-)}_{A}
\label{KS-}
\ee
where $\eta^{(-)}_{A}$ is a constant spinor. 

\section{BPS states in the $\cN=4$ SYM} 
\label{BPSN=4SYM}
In this section we perform a systematic analysis on 
BPS states of $\cN=4$ SYM on $\mbf R\times \mbf S^3$. 
That is we study a condition of the fields such that the supersymmetry variation 
given by \eqref{susy} vanishes for a certain Killing spinor. 
In this paper we consider a situation where the gaugino fields vanish. 
In this case we have only to study conditions of the bosonic fields for the supersymmetry
transformation of gaugino to vanish.  
By using the Killing spinor equation \eqref{KSequation}
the supersymmetry variation of gaugino can be written as
\beal{
\Delta_{\epsilon^{(\pm)}} \lambda_{A} 
=&\half F_{{\mu}{\nu}}\gamma^{{\mu}{\nu}} \epsilon^{(\pm)}_{A} -2 i[X_{AB}, X^{BC}] \epsilon^{(\pm)}_{C} + (-2 D_0 X_{AB} \pm X_{AB} {2i \over r} ) \gamma^t (\epsilon^{(\pm)}_B)^*
- 2 D_i X_{AB} \gamma^i (\epsilon^{(\pm)}_B)^*
\label{deltalambda}
}
where $i=\theta, \phi, \psi$. 

There are two kinds of BPS states on the sphere: one is spherically symmetric, the other is not. 
This separation is achieved automatically by studying whether a preserved Killing spinor of a BPS state is projected or not by a certain projective operator, which breaks spherical symmetry.
In this paper we perform projection for a constant spinor in Killing spinors \eqref{KS+}, \eqref{KS-} by using $\gamma_0$.

\subsection{Spherically symmetric BPS states }
\label{sphericalsym}

In this subsection we study spherically symmetric BPS solutions for warm-up.  
For this purpose we study a BPS condition which preserves 
at least one Killing spinor without any projection. 
As such a preserved Killing spinor 
we first choose $\epsilon^{(+)}_{\mathfrak 4}$,  
where $\mathfrak 4 = 1,2,3,4$. 
We use German letters $\mf1, \mf2, \mf3, \mathfrak 4$ as symbols representing $1,2,3,4$ 
in a way that there exists a permutation $\sigma$ such that $\mf i = \sigma(i)$, 
where $\mf i =\mf1, \mf2, \mf3, \mathfrak 4, ~ i = 1,2,3,4$. 
For the supersymmetry variation of gaugino to vanish with nonzero $\epsilon^{(+)}_{\mathfrak 4}$, 
the fields are required to satisfy
\beal{
F_{{\mu}{\nu}}=0, \quad [X_{AB}, X^{B\mf4}]=0, \quad
- 2 D_0 X_{A \mathfrak 4} + X_{A \mathfrak 4} {2i \over r} = 0, \quad 
D_i X_{A\mathfrak 4} =0
\label{bps1/2}
}
for $A=\mf1, \mf2, \mf3, \mf4$, since  $\gamma^{\mu} \epsilon^{(+)}_{\mf4}$ and $( \gamma^{\mu} \epsilon^{(+)}_\mf4)^*$ are linearly independent for all $\mu$.
This is the 1/8 BPS condition which preserves a Killing spinor $\epsilon^{(+)}_{\mathfrak 4}$. 
As asserted, the last equation in \eqref{bps1/2} constrains BPS states to be spherically symmetric.

Let us study this BPS condition when the fields take values in Cartan subalgebra of the gauge group.  
In this case these become 
\beal{
F_{{\mu}{\nu}}=0, \quad 2 \partial_0 X_{A \mathfrak 4} = X_{A \mathfrak 4} {2i \over r}, \quad 
\partial_i X_{A\mathfrak 4} =0.
\label{bps1/2cartan}
}
This can be easily solved as 
\be
X_{A\mf4} = x_{A\mf 4}^{(+)} e^{i {t\over r}}
\label{bps1/2cartansolution1}
\ee
where $x_{A\mf 4}$ are integral constants valued in complex number. 
Other components are determined by using the relation \eqref{Xconstraint} as  
\beal{
X_{\mf 1 \mf 2} = X_{\mf 3\mf 4}^\dagger= x^{(+)} _{\mf 3\mf4}{}^\dagger e^{-i {t\over r}}, \quad 
X_{\mf 1 \mf 3} =- X_{\mf 2\mf 4}^\dagger=- x_{\mf 2\mf4}^{(+)} {}^\dagger e^{-i {t\over r}}, \quad 
X_{\mf 2 \mf 3} = -X_{\mf 1\mf 4}^\dagger=- x_{\mf 1\mf4}^{(+)} {}^\dagger e^{-i {t\over r}}. \quad 
}
This is a general 1/8 BPS solution which preserves a Killing spinor $\epsilon^{(+)}_{\mathfrak 4}$. 

Let us study a case for this BPS solution to preserve another Killing spinor. 
One will soon notice that $\epsilon^{(-)}_{\mf4}$ is broken unless all fields are trivial. 
So let us set it to zero to study a nontrivial BPS solution. 
Under the BPS condition \eqref{bps1/2} the supersymmetry variation of gaugino becomes
\beal{
\Delta_{\epsilon^{(\pm)}} \lambda_{\mf1} 
=& (X_{\mf1\mf2} {2i \over r} \pm X_{\mf1\mf2} {2i \over r} ) \gamma^t (\epsilon^{(\pm)}_{\mf2})^* +(X_{\mf1\mf3} {2i \over r}  \pm X_{\mf1\mf3} {2i \over r} ) \gamma^t (\epsilon^{(\pm)}_{\mf3})^*, \nn
\Delta_{\epsilon^{(\pm)}} \lambda_{\mf2} 
=& (X_{\mf2\mf1} {2i \over r}\pm X_{\mf2\mf1} {2i \over r} ) \gamma^t (\epsilon^{(\pm)}_{\mf1})^* +(X_{\mf2\mf3} {2i \over r} \pm X_{\mf2\mf3} {2i \over r} ) \gamma^t (\epsilon^{(\pm)}_{\mf3})^*, \nn
\Delta_{\epsilon^{(\pm)}} \lambda_{\mf3} 
=& ( X_{\mf3\mf1} {2i \over r} \pm X_{\mf3\mf1} {2i \over r} ) \gamma^t (\epsilon^{(\pm)}_{\mf1})^*
 +(X_{\mf3\mf2} {2i \over r} \pm X_{\mf3\mf2} {2i \over r} ) \gamma^t (\epsilon^{(\pm)}_{\mf2})^*,\nn
 \Delta_{\epsilon^{(\pm)}} \lambda_{\mf4} 
=& (- X_{\mf4\mf1} {2i \over r} \pm X_{\mf4\mf1} {2i \over r} ) \gamma^t (\epsilon^{(\pm)}_{\mf1})^*
 +(- X_{\mf4\mf2} {2i \over r} \pm X_{\mf4\mf2} {2i \over r} ) \gamma^t (\epsilon^{(\pm)}_{\mf2})^*+(-X_{\mf4\mf3} {2i \over r}  \pm X_{\mf4\mf3} {2i \over r} ) \gamma^t (\epsilon^{(\pm)}_{\mf3})^*. 
}
In order for these to vanish, it is required to satisfy 
\beal{
&(\epsilon^{(+)}_{\mf1})^* =0\quad \text{or} \quad
(\epsilon^{(+)}_{\mf1})^* \not=0, X_{\mf 3\mf4} = X_{\mf2\mf4}=0,\\ 
&(\epsilon^{(+)}_{\mf2})^* =0\quad \text{or} \quad
(\epsilon^{(+)}_{\mf2})^* \not=0, X_{\mf 3\mf4} = X_{\mf1\mf4}=0,\\ 
&(\epsilon^{(+)}_{\mf3})^* =0\quad \text{or} \quad
(\epsilon^{(+)}_{\mf3})^* \not=0, X_{\mf 2\mf4} = X_{\mf1\mf4}=0,\\ 
&(\epsilon^{(-)}_{\mf1})^* =0\quad \text{or} \quad
(\epsilon^{(-)}_{\mf1})^* \not=0, X_{\mf4\mf1} =0,  \\ 
&(\epsilon^{(-)}_{\mf2})^* =0\quad \text{or} \quad
(\epsilon^{(-)}_{\mf2})^* \not=0, X_{\mf4\mf2} =0, \\ 
&(\epsilon^{(-)}_{\mf3})^* =0\quad \text{or} \quad
(\epsilon^{(-)}_{\mf3})^* \not=0, X_{\mf4\mf3} =0,
}
where we used the relation \eqref{Xconstraint}. 
Therefore the 1/8 BPS solution has enhanced supersymmetry in the following situation. 
\begin{enumerate}
 \item One complex scalar field is trivial: $X_{\mf1\mf4}=0$. 
 In this case, the Killing spinor $\epsilon^{(-)}_{\mf1}$ is also preserved and the solution \eqref{bps1/2cartansolution1} becomes 1/4 BPS.
 \item Two scalar fields are trivial: $X_{\mf1\mf4}=X_{\mf2\mf4}=0$. 
 In this case, the Killing spinors $\epsilon^{(-)}_{\mf1}, \epsilon^{(-)}_{\mf2}, \epsilon^{(+)}_{\mf3}$ are also conserved and the solution \eqref{bps1/2cartansolution1} becomes 1/2 BPS.
\end{enumerate}

In the same way we can obtain a BPS condition and solution which preserve $\epsilon^{(-)}_{\mf4}$. 
The BPS condition to conserve $\epsilon^{(-)}_{\mf4}$ is 
\beal{
F_{ {\mu} {\nu}}=0, \quad [X_{AB}, X^{B\mf4}]=0, \quad
2 D_0 X_{A \mathfrak 4} - X_{A \mathfrak 4} {2i \over r} = 0, \quad 
D_i X_{A\mathfrak 4} =0.
\label{bps1/2-}
}
When the fields take values in Cartan subalgebra of the gauge group, 
this reduces to 
\beal{
F_{ {\mu} {\nu}} =0, \quad 
2 \partial_0 X_{A\mf4} + X_{A\mf4} {2i \over r} = 0, \quad 
\partial_i X_{A\mf4} =0.
\label{bps1/2cartan-}
}
The equations of the complex scalar field can be easily solved as
\be
X_{A\mf4} = x^{(-)} _{A\mf 4} e^{- i {t\over r}}.
\label{bpssolution2}
\ee
This is a general 1/8 BPS solution which preserves $\epsilon^{(-)}_{\mf4}$. 
This BPS solution has enhanced supersymmetry 
in the following situation. 
\begin{enumerate}
 \item One complex scalar field is trivial: $X_{\mf1\mf4}=0$. 
 In this case, the Killing spinor $\epsilon^{(+)}_{\mf1}$ is also conserved, and  
the solution becomes 1/4 BPS.
 \item Two scalar fields are trivial: $X_{\mf1\mf4}=X_{\mf2\mf4}=0$. 
 In this case, the Killing spinors $\epsilon^{(+)}_{\mf1}, \epsilon^{(+)}_{\mf2}, \epsilon^{(-)}_{\mf3}$ are also unbroken, and the solution becomes 1/2 BPS.
\end{enumerate}

We summarize the result in the following table. 
\begin{table}[htbp]
\begin{center}
\begin{tabular}{c|c|c}
\hline
BPS solutions. &  Preserved Killing spinors. & Number of SUSY.  \\ \hline \hline
$X_{A\mf4} = x^{(\pm)} _{A\mf 4} e^{\pm i {t\over r}}$ for $A=\mf1,\mf2,\mf3$. & $\epsilon^{(\pm)}_{\mf4}$.   &  $4$  \\
 &  & (${1 \over 8}$ BPS) \\
\hline
$X_{A\mf4} = x^{(\pm)} _{A\mf 4} e^{\pm i {t\over r}}$ for $A=\mf2,\mf3$, & $\epsilon^{(\mp)}_{\mf1},\epsilon^{(\pm)}_{\mf4}$.   &  $8$  \\
$X_{\mf1\mf4} =0$. &  & (${1 \over 4}$ BPS) \\
\hline
$X_{\mf3\mf4} = x^{(\pm)} _{\mf3\mf 4} e^{\pm i {t\over r}}$,   & $\epsilon^{(\mp)}_{\mf1},\epsilon^{(\mp)}_{\mf2},\epsilon^{(\pm)}_{\mf3},\epsilon^{(\pm)}_{\mf4}$.   &  $16$  \\
$X_{\mf1\mf4} =X_{\mf2\mf4} =0$.  &  & (${1 \over 2}$ BPS) \\
\hline
$X_{\mf1\mf4} =X_{\mf2\mf4}=X_{\mf3\mf4} =0$.  & $\epsilon^{(\pm)}_{\mf1},\epsilon^{(\pm)}_{\mf2},\epsilon^{(\pm)}_{\mf3},\epsilon^{(\pm)}_{\mf4}$.   &  $32$  \\
 &  & (Unique vacuum) \\
\hline
\end{tabular}
\caption{We list spherically symmetric BPS solutions, preserved Killing spinors and its number.  
Signature in multi-column has to be chosen with order respected except the last column.   
}
\label{sphericalBPS}
\end{center}
\end{table}

Let us compute conserved charges \eqref{energy}, \eqref{momenta}, \eqref{Rcharge} 
under the BPS conditions. 
\beal{
H=& {1\over g^2} \int d^3\Omega~ {4 \over r^2} \Tr |X_{A\mf4}|^2, \\
P_{i} =& 0, \\
R^{\mf4}{}_{\mf4} 
=&  \int d^3\Omega~   {2\over g^2}  {\mp 1\over r} \Tr | X_{A \mf4} |^2, 
}
where the upper sign is for the BPS condition given by \eqref{bps1/2} and the lower for that of \eqref{bps1/2-}. 
We can show that $\sum_{A=1}^4 R^A{}_A=0$.
We can easily see the BPS relation of conserved charge as  
\be
r H =\mp 2 R^{\mf4}{}_{\mf4} =\pm 2 \sum_{A=\mf1}^{\mf3} R^A{}_A. 
\ee

\subsection{Non-spherical BPS states }
\label{nonsphericalBPS}
\subsubsection{BPS states preserving $\epsilon^{(+)}_\mf4$} 
\label{1/16BPSstates+}

In this subsection we study supersymmetric states with angular momenta. 
To this end we study supersymmetric states which preserves a Killing spinor projected by a certain operator 
which breaks spherical symmetry. 
We first choose $\epsilon^{(+)}_{\mf4}$ as such a preserved Killing spinor.  
We carry out projection for a constant spinor $\eta^{(+)}_{\mf4}$ such that
\beal{
&\gamma_0 \eta_{\mf4}^{(+)}= i w^{(+)}\eta_{\mf4}^{(+)} 
\label{projection+}
}
where $w^{(+)}= \pm 1$.%
\footnote{ 
In terms of a Killing spinor the projection is given by 
$
\cP \epsilon_{\mf4}^{(+)}  = i w^{(+)} \epsilon_{\mf4}^{(+)} 
$
where $\cP = U \gamma_0 U^{-1}$ with $\epsilon_{\mf4}^{(+)} = U \eta_{\mf4}^{(+)}$. 
}
For this projected constant spinor, 
the Killing spinor $\epsilon_{\mf4}^{(+)}$ given by \eqref{KS+} is evaluated as 
\beal{
\epsilon^{(+)}_{\mf4} =&  e^{{i \over 2r} t} e^{ i {w^{(+)} \over 2}  \psi} \eta^{(+)}_{\mf4}.
\label{KS+projected}
}
Let us fix charges of this Killing spinor. 
The energy, the angular momentum of $\phi$ direction, and that of $\psi$ direction 
are evaluated as eigenvalues of the operators $H:=i\partial_t$, $J_\phi:=\hat L_3=-i\partial_\phi$, 
and $J_\psi:=\hat R_3=i\partial_\psi$, respectively, 
where $\hat L_i, \hat R_i$ are defined in Appendix \ref{s3}. 
The charge assignments are summarized in Table \ref{charge+}. 
\begin{table}[h]
\begin{center}
\begin{tabular}{cccccccccc}
\hline
  &  $H$ & $J_\phi$  & $J_\psi$ & $R^{\mf1}{}_{\mf1}$& $R^{\mf2}{}_{\mf2}$& $R^{\mf3}{}_{\mf3}$  &$ R^{\mf4}{}_{\mf4}$\\
\hline
\hline
$\epsilon^{(+)}_{\mf4}$ & $-{1\over 2r} $ & $0$ & $ - {w^{(+)} \over 2}$ & $-{1 \over 4}$ & $-{1 \over 4}$ & $-{1 \over 4}$ & ${3 \over 4}$ \\
\hline
\end{tabular}
\caption{Charges of the Killing spinor $\epsilon^{(+)}_{\mf4}$ are shown. }
\label{charge+}
\end{center}
\end{table}
Therefore the BPS condition of charges associated with the Killing spinor $\epsilon^{(+)}_{\mf4}$ is 
\be
r H = - 2 R^{\mf4}{}_{\mf4} + 2  w^{(+)} J_\psi. 
\label{bpscharge1}
\ee
Note that the relative factors match that obtained from $psu(2,2|4)$ superconformal algebra
in the standard normalization \cite{Dolan:2002zh,Kinney:2005ej}. 

In order to study a BPS condition preserving $\epsilon_{\mf4}^{(+)}$, 
it is convenient to divide the supersymmetric transformation of gaugino into two parts such that 
\beal{
\Delta_{\epsilon^{(\pm)}} \lambda_{C} =&\Delta_{{(\pm)}} \lambda_{C} + \bar\Delta_{{(\pm)}} \lambda_{C}
\label{decomposition}
}
where 
\beal{ 
\Delta_{{(\pm)}} \lambda_{A}
=& \half F_{ {\mu} {\nu}}\gamma^{ {\mu} {\nu}} \epsilon^{(\pm)}_{A} -2 i[X_{AB}, X^{BC}] \epsilon^{(\pm)}_{C}, 
\label{dec1}\\
\bar\Delta_{{(\pm)}} \lambda_{A}
=&  (- 2 D_0 X_{AB} \pm X_{AB} {2i \over r} ) \gamma^t (\epsilon^{(\pm)}_B)^* - 2 D_i X_{AB} \gamma^i (\epsilon^{(\pm)}_B)^*.
\label{dec2}
}
Let us extract the terms containing $\eta_{\mf4}^{(+)}$ from these. 
By using \eqref{projection+} and \eqref{KS+projected}. 
we find 
\beal{
\Delta_{{(+)}} \lambda_{A}
=&e^{ {i\mu \over 4} t}e^{ i {w^{(+)} \over 2}  \psi} \bigg[iw^{(+)} \delta^\mf4_A( - F^{01} + i F^{23} + iw^{(+)} F^{02} - w^{(+)} F^{13} )\gamma_1 \eta^{(+)}_{\mf4} \nn
&+ \{ iw^{(+)} \delta^\mf4_A (i F^{03} + F^{12} ) +  ( -2 i[X_{AB}, X^{B\mf4}] ) \}\eta^{(+)}_{\mf4} \bigg] + \cdots \\
\bar\Delta_{{(+)}} \lambda_{A}
=&e^{- {i\mu \over 4} t}e^{ i {-w^{(+)} \over 2}  \psi} \bigg[ 
\( (- 2 D_0 X_{A\mf4} + X_{A\mf4}{2i \over r})(iw^{(+)}) - 2i D_3 X_{A\mf4} \) (\eta^{(+)}_\mf4)^* \nn
& +(- 2 D_1 X_{A\mf4} - 2 i D_2 X_{A\mf4} w^{(+)}) \gamma^1(\eta^{(+)}_\mf4)^*  \bigg]   +\cdots
}
where the ellipses represent the other terms which do not contain $\eta^{(+)}_{\mf4}$.
In order for $\Delta_{\epsilon^{(\pm)}} \lambda_{C}$ to vanish with $\eta^{(+)}_{\mf4}$ nonzero, 
it is necessary to satisfy
 \beal{
 &  - F^{01} + i F^{23} + iw^{(+)} F^{02} - w^{(+)} F^{13}  =0,\nn
 &iw^{(+)} \delta^{\mf4}_A ( i F^{03} + F^{12}) -2 i[X_{AB}, X^{B\mf4}] =0,\nn
& (- 2 D_0 X_{A\mf4} +X_{A\mf4}{2i \over r})(iw^{(+)})- 2i D_3 X_{A\mf4}=0, \nn
&- 2 D_1 X_{A\mf4} - 2 i D_2 X_{A\mf4} w^{(+)}=0, \notag
 }
 for arbitrary $A$, since $\eta^{(+)}_{\mf4}, \gamma_1 \eta^{(+)}_{\mf4}, ( \eta^{(+)}_{\mf4})^*, \gamma_1 (\eta^{(+)}_{\mf4})^*$ are linearly independent under the projection \eqref{projection+}. 
By using the fact that all the components of $F^{\mu\nu}$ are real, 
these can be simplified as 
 \begin{equation}\begin{split}
 & F^{01}= - w^{(+)} F^{13}, \quad F^{23} =- w^{(+)} F^{02}, \\
 &w^{(+)} \delta^{\mf4}_A ( i F^{03} + F^{12}) -2 [X_{AB}, X^{B\mf4}] =0,\\
& (D_0 -{i \over r}- w^{(+)} D_3) X_{A\mf4}=0, \\
& ( D_1 + i w^{(+)} D_2) X_{A\mf4} =0.
\label{bps+general}
\end{split}\end{equation}
This is the 1/16 BPS condition preserving $\epsilon^{(+)}_{\mf4}$ projected by \eqref{projection+}. 
This result is essentially the same as 1/16 BPS equations derived in \cite{Grant:2008sk}, (4.9) and (4.10),  in a different way 
by making the energy density complete square and looking for configurations to saturate the Bogomol'nyi bound. 

Let us solve these BPS conditions when the fields are valued in Cartan subalgebra of a gauge group.  
In this case these BPS equations boil down to 
\be
F^{12} =F^{03}=0, \quad  
F^{13} =- w^{(+)} F^{01}, \quad  
F^{02} =- w^{(+)} F^{23}, \quad  
\label{bpsF+}
\ee
and 
\be
(\partial_0 -{i \over r}- w^{(+)} \partial_3) X_{A\mf4}=0, \quad ( \partial_1 + i w^{(+)} \partial_2) X_{A\mf4} =0 
\label{bpsX+}
\ee
for $A = \mf1, \mf2, \mf3$. 
Note that by restricting our interest to the Cartan part 
the gauge sector and the matter one decouple so that we can solve each sector independently.%
\footnote{ 
This is not the case in the non-abelian sector \cite{Grant:2008sk}.}

First let us solve the matter BPS equation.
The equation of motion of the matter fields \eqref{eomX} suggests 
that solutions thereof are expanded by $\mbf S^3$ scalar spherical harmonics. 
We denote the scalar spherical harmonics by $Y_{s, l_3, r_3}$ 
with non-negative half integer $s$ and two half integers $l_3, r_3$ whose moduli are bounded above by $s$.
These quantum numbers are Cartan charges of the representation of the spherical harmonics of $su(2)_L\times su(2)_R$ acting on $\mbf S^3$. 
\beal{
&\hat L^2 Y_{s, l_3, r_3} = \hat R^2 Y_{s, l_3, r_3} = s(s+1) Y_{s, l_3, r_3}, \quad \\
&\hat L_3 Y_{s, l_3, r_3} = l_3 Y_{s, l_3, r_3}, \quad 
\hat R_3 Y_{s, l_3, r_3} = r_3 Y_{s, l_3, r_3},
}
where $\hat L_i, \hat R_i$ are differential operators generating $su(2)_L\times su(2)_R$ algebra.  
See Appendix \ref{s3} for more details. 
By using these operators the second equation in \eqref{bpsX+} can be written as 
\be
\hat R_{w^{(+)}} X_{A\mf4} =0
\ee
where $\hat R_{w^{(+)}} = \hat R_{\pm}$ when $w^{(+)} = \pm1$ respectively, 
and $\hat R_\pm$ is given by \eqref{LRpm}. 
Therefore BPS solutions are expanded by the spherical harmonics of highest (or lowest) weight of $su(2)_R$ for $w^{(+)}=+1$ (or $-1$).
\be
X_{A\mf4} = \sum_{s\geq0 , |l_3|\leq s} x_{A\mf4}^{s, l_3}(t) Y_{s, l_3, w^{(+)} s}(\theta,\phi,\psi)
\ee
where $x_{A\mf4}^{s, l_3}(t)$ are unknown functions of time, 
which is determined from the first equation in \eqref{bpsX+}. 
Plugging the above into the first equation in \eqref{bpsX+} gives 
\be
\partial_0 x_{A\mf4}^{s, l_3}(t)=  { i\over r}x_{A\mf4}^{s, l_3}(t) - {2 \over r} w^{(+)} (-i w^{(+)} s) x_{A\mf4}^{s, l_3}(t).
\ee
This equation is easily solved as 
\be
x_{A\mf4}^{s, l_3}(t) = x_{A\mf4}^{s, l_3} e^{i {2s+1\over r}t}
\ee
where $x_{A\mf4}^{s, l_3}$ is an integral constant valued in complex number.%
\footnote{\label{abusage}
More rigorously or mathematically speaking, $x_{A\mf4}^{s, l_3}$ is a constant valued in Cartan subalgebra of a gauge group over the complex number. But we often abuse such a variable as a coefficient of a Cartan generator by which the variable is expanded. 
} 
As a result we obtain 
\be
X_{A\mf4} = \sum_{s\geq0 , |l_3|\leq s} x_{A\mf4}^{s, l_3} e^{i {2s+1\over r}t}Y_{s, l_3, w^{(+)} s}(\theta,\phi,\psi).
\label{1/16BPS+Xsol}
\ee
This is a general $1/16$ BPS solution of the complex scalar fields preserving the Killing spinor $\epsilon^{(+)}_\mf4$ with projection \eqref{projection+}. 
Note that this satisfies the equation of motion of the scalar field \eqref{eomX}. 

Let us move on to determining BPS solutions of gauge sector. 
For this purpose we substitute the BPS condition \eqref{bpsF+} into the equations of motion of the field strength \eqref{eomF} and one of the Bianchi identities, which are in our situation given by 
$
\nabla_{\mu}F^{\mu\nu}=0,  
$
$
\nabla_1 F^{23}+\nabla_2 F^{31}+\nabla_3 F^{12} =0, 
$
respectively. 
Under the BPS condition \eqref{bpsF+} 
the equations of motion reduce to the following three equations  
\beal{ 
&\partial_{t} F^{01} +w^{(+)} \partial_{3} F^{01}=0, 
\label{1/16BPS+D0F01}\\
&\partial_{t} F^{23} +w^{(+)}  \partial_{3} F^{32} =0, 
\label{1/16BPS+D0F23}\\
&\partial_{1}(  w^{(+)} F^{01}) - \partial_{2} F^{23} +{2\over r} \cot\theta( w^{(+)} F^{01})  =0,  
\label{1/16BPS+D3F23}
}
and the Bianchi identity becomes 
\be
\partial_{1}(\sin\theta  F^{23}) + w^{(+)}\partial_{2}( \sin\theta F^{01}) =0.
\label{bianchibps}
\ee
After multiplying $\sin\theta$ to both sides in \eqref{1/16BPS+D3F23}, we can write it as 
\be
\partial_{1}(\sin\theta  F^{01}) - w^{(+)}\partial_{2}( \sin\theta F^{23}) =0.
\label{1/16BPS+D3F23'} 
\ee
A general solution of \eqref{bianchibps} and \eqref{1/16BPS+D3F23'} is given by 
\begin{equation}\begin{split}
\sin\theta  F^{01}= &\sum_{s\geq\half, |l_3|\leq s} (B^{s, l_3} (t)\text{Re}[Y_{s, l_3, w^{(+)} s}(\theta,\phi,\psi)] - A^{s, l_3} (t)\text{Im}[Y_{s, l_3, w^{(+)} s}(\theta,\phi,\psi)] ) \\
\sin\theta  F^{23}= & \sum_{s\geq\half, |l_3|\leq s} (A^{s, l_3} (t)\text{Re}[Y_{s, l_3, w^{(+)} s}(\theta,\phi,\psi)]+B^{s, l_3} (t)\text{Im}[Y_{s, l_3, w^{(+)} s}(\theta,\phi,\psi)] )  
\label{1/16BPS-D3F23sol}
\end{split}\end{equation}
where $A^{s, l_3} (t), B^{s, l_3} (t)$ are unknown functions of time. 
The reason is as follows. 
First we recall that $Y_{s, l_3, w^{(+)} s}$ is annihilated by the operator $\hat R_{w^{(+)}}$. 
By dividing $Y_{s, l_3, w^{(+)} s}$ into the real and imaginary part, $Y_{s, l_3, w^{(+)} s}=u+iv$, 
we can rewrite $\hat R_{w^{(+)}} Y_{s, l_3, w^{(+)} s}=0$ as 
\beal{
&\partial_1 u -  w^{(+)} \partial_2 v  =0, \quad
\partial_1 v +  w^{(+)} \partial_2 u  =0. \quad
}
This is equivalent to 
\beal{
&\partial_1 (Au+Bv) +w^{(+)} \partial_2 (Bu-Av) =0, \quad
\partial_1 (Bu-Av) - w^{(+)} \partial_2 (Au+Bv) =0
}
where $A, B$ are arbitrary constants independent of $\theta, \phi, \psi$. 
This implies that the equations \eqref{bianchibps} and \eqref{1/16BPS+D3F23'} are solved by 
$\sin\theta F^{23} = Au+Bv, ~ \sin\theta F^{01}= Bu-Av$. 
Superposing all modes labeled by $l, l_3$ gives a general solution \eqref{1/16BPS-D3F23sol}, 
as asserted. 

The time dependence of the field strength is  determined by \eqref{1/16BPS+D0F01}, \eqref{1/16BPS+D0F23}.  
The result is 
\beal{
\sin\theta  F^{01}
= &  \sum_{s\geq\half , |l_3|\leq s} a^{(+)}_{s, l_3} \text{Re}[Y_{s, l_3, w^{(+)} s}(\theta,\phi,\psi-w^{(+)} {2 \over r} t+\alpha^{(+)}_{s, l_3})] \\
\sin\theta  F^{23}= & \sum_{s\geq\half, |l_3|\leq s} a^{(+)}_{s, l_3}\text{Im}[Y_{s, l_3, w^{(+)} s}(\theta,\phi,\psi-w^{(+)} {2 \over r} t+\alpha^{(+)}_{s, l_3})] 
}
where $a^{(+)}_{s, l_3}, \alpha^{(+)}_{s, l_3}$ are integral constants taking real values with the range
$
a^{(+)}_{s, l_3} \geq 0, ~ 0\leq \alpha^{(+)}_{s, l_3}< 2\pi.
$
Plugging the explicit expression of the spherical harmonics given by \eqref{su2Rhighest} into these, 
we find the general 1/16 BPS solution of the field strength as
\beal{
F^{01}
= &  \sum_{s\geq\half , |l_3|\leq s} a^{(+)}_{s, l_3} c_{s,l_3} \tan^{w^{(+)} l_3}\left(\frac{\theta }{2}\right) \sin ^{s-1}\theta \cos{(l_3\phi  +s({2 \over r} t-w^{(+)}  \psi)+\alpha^{(+)}_{s, l_3})} 
\label{1/16BPS+F01sol}\\
F^{23}= & \sum_{s\geq\half , |l_3|\leq s} a^{(+)}_{s, l_3} c_{s,l_3} \tan^{w^{(+)} l_3}\left(\frac{\theta }{2}\right) \sin ^{s-1}\theta \sin{(l_3\phi +s({2 \over r} t-w^{(+)}  \psi)+\alpha^{(+)}_{s, l_3})} 
\label{1/16BPS+F23sol}
}
where $c_{s,l_3}$ is given by \eqref{csl3}.%
\footnote{
Here we exclude the mode of $s=0$, which formally satisfies the $1/16$ BPS equations but turns out to be non-normalizable.   
}  
Other components of the field strength are determined by \eqref{bpsF+}. 

Let us enphasize that it is possible to turn on BPS excitation for the gauge field 
independently from the matter sector in the abelian sector. 
This behavior is different from the non-abelian sector \cite{Grant:2008sk}.%
\footnote{The author would like to thank S.Kim for pointing this out. }
We also note that it is not required to use the vector spherical harmonics 
to solve the BPS equation for the gauge field. 

The moduli space of the $1/16$ BPS solution is given by
\be
x_{\mf1\mf4}^{s, l_3},x_{\mf2\mf4}^{s, l_3},x_{\mf3\mf4}^{s, l_3} \in \mbf C, 
~ a^{(+)}_{s, l_3} \geq 0, ~ 0\leq \alpha^{(+)}_{s, l_3}< 2\pi
\label{modulispace+}
\ee 
for $s\geq0, |l_3|\leq s$ for $x_{A\mf4}^{s, l_3}$, $s\geq\half, |l_3|\leq s$ for $a^{(+)}_{s, l_3}$. 
Note that we do not take into account the flux quantization condition here and hereafter,
which would make the moduli space quantized in a certain manner. 

Let us compute conserved charges \eqref{energy}, \eqref{momenta} and \eqref{Rcharge}. 
We first simply them by using the BPS solution \eqref{bpsF+}, \eqref{bpsX+}.
As a result the conserved charges are written in terms of $X_{A\mf4}, \partial_2 X_{A\mf4}, \partial_3 X_{A\mf4}, F^{01}, F^{23}$.
We present results by separating the matter sector and gauge sector.  
The conserved charges in the matter sector are simplified as 
\beal{
H|_{F_{\mu\nu}=0}
=&\int d^3\Omega~ {4\over g^2}   \Tr\bigg( {1\over r^2} |X_{A\mf4}|^2 - w^{(+)} { i\over r} X_{A\mf4} \partial_3 X^{A\mf4}+  | \partial_2 X_{A\mf4}|^2 + | \partial_3 X_{A\mf4}|^2   \bigg), \\
J_{\theta}|_{F_{\mu\nu}=0} =&0,  \\
J_{\phi}|_{F_{\mu\nu}=0} =&  \int d^3\Omega~   {2\over g^2}  \Tr \(- (   { i\over r} X_{A\mf4} - w^{(+)} \partial_3 X_{A\mf4})  \partial_\phi X^{A\mf4} +(c.c.) \),  \\
J_{\psi}|_{F_{\mu\nu}=0} =& \int d^3\Omega~   {2\over g^2} \Tr \(- (   {i\over r} X_{A\mf4} - w^{(+)} \partial_3 X_{A\mf4}) \partial_\psi X^{A\mf4} +(c.c.) \),  \\
R^{\mf4}{}_{\mf4}|_{F_{\mu\nu}=0}  =& i  \int d^3\Omega~   {2\over g^2}\Tr \({ i\over r} |X_{\mf4B}|^2 - w^{(+)} X^{\mf4B} \partial_3 X_{\mf4B} \),
}
where $(c.c.)$ means the complex conjugation of the first term. 
The gauge field part is the following. 
\beal{
H|_{X^{AB}=0} =&   \int d^3\Omega~  {1\over g^2} \Tr\((F^{01})^2+ (F^{23})^2 \), \\
J_{\theta}|_{X^{AB}=0} =& 0, \\
J_{\phi}|_{X^{AB}=0}
=&w^{(+)}   {r \over 2g^2}  \int d^3\Omega~ \cos\theta \Tr \((F^{01})^2  + ( F^{23})^2 \), \\
J_{\psi}|_{X^{AB}=0} 
=&w^{(+)}  {r \over 2g^2} \int d^3\Omega~ \Tr \((F^{01})^2  + ( F^{23})^2 \), \\
R^{\mf4}{}_{\mf4}|_{X^{AB}=0}   =&0. 
}
Note that at this stage we see the BPS relation of charges given by \eqref{bpscharge1} in the gauge sector. 

Let us compute conserved charges using the BPS solution \eqref{1/16BPS+Xsol}, \eqref{1/16BPS+F01sol}, \eqref{1/16BPS+F23sol}.
The matter part is the following.%
\footnote{
For this computation we used formulas such that 
\beal{
 \int d^3\Omega~   {1\over\sin^2\theta} |Y_{s, l_3, w^{(+)} s}|^2=& \frac{s (1+2 s)}{2 \left(-l_3^2+ s^2\right)}, \\
 \int d^3\Omega~    {\cot\theta\over\sin\theta}  |Y_{s, l_3, w^{(+)} s}|^2=&\frac{w^{(+)}l_3(1+2 s)}{2 (l_3^2- s^2)},\\
 \int d^3\Omega~  \cot^2\theta |Y_{s, l_3, w^{(+)} s}|^2=&-\frac{2 l_3^2+ s}{2( l_3^2-s^2)},
}
 where $|l_3|<s$. In the main text, we also used these formulas formally at $|l_3|=s$. 
} 
\beal{
H|_{F_{\mu\nu}=0}
=& {8 \over r^2 g^2}\sum_{s\geq0 , |l_3|\leq s}  (2s+1)^2  \Tr |x_{A\mf4}^{s, l_3}|^2, \\
J_{\phi} |_{F_{\mu\nu}=0}
=& {- 4\over r g^2}  \sum_{s\geq0 , |l_3|\leq s}  (2s +1 ) l_3 \Tr |x_{A\mf4}^{s, l_3}|^2,  \\
J_{\psi} |_{F_{\mu\nu}=0}
=& {4\over r g^2} w^{(+)}  \sum_{s\geq0 , |l_3|\leq s}  (2s+1) s \Tr |x_{A\mf4}^{s, l_3}|^2,\\
R^{\mf4}{}_{\mf4}|_{F_{\mu\nu}=0} =& -{2\over r g^2}  \sum_{s\geq0 , |l_3|\leq s} (2s +1 ) \Tr |x_{A\mf4}^{s, l_3} |^2. 
}
In the matter sector we also find the BPS relation given by \eqref{bpscharge1}.  

The gauge field part is computed as follows. 
\beal{
& H|_{X_{AB}=0} =w^{(+)} {2\over r} J_{\psi}|_{X_{AB}=0}
=  {1\over g^2}\sum_{s\geq\half , |l_3|\leq s}    \frac{s (2 s +1)}{2 \left(s^2-l_3^2\right)} \Tr(a^{(+)}_{s, l_3})^2, \\
&J_{\phi}|_{X_{AB}=0}= {r \over 2 g^2}  \sum_{s\geq\half, |l_3|\leq s}   \frac{ l_3(2 s +1)}{2 (l_3^2- s^2)}
 \Tr (a^{(+)}_{s, l_3})^2.  
 }
 Note that the energy and momenta become divergent when the modes with $|l_3|=s$ are nonzero.  

\subsubsection{BPS states preserving $\epsilon^{(-)}_\mf1$} 
\label{1/16BPSstates-}

Next we study BPS states preserving a Killing spinor $\epsilon^{(-)}_\mf1$ whose constant spinor $\eta^{(-)}_{\mf1}$ is projected in such a way that 
\beal{
&\gamma_0 \eta_{\mf1}^{(-)}= i w^{(-)}\eta_{\mf1}^{(-)}
\label{projection-}
}
where $w^{(-)}= \pm 1$. Under this projection 
the Killing spinor $\epsilon^{(-)}_\mf1$ given by \eqref{KS-} is evaluated as
\beal{
\epsilon^{(-)}_{\mf1} =&  e^{- {i \over 2r} t}e^{- iw^{(-)} \phi \over 2} (c_\theta - w^{(-)} \gamma_1 s_\theta)  \eta^{(-)}_{\mf1} 
}
where we used notation such that 
\be
c_\theta=\cos{\theta\over2}, \quad s_\theta=\sin{\theta\over2}.
\ee

We determine charges of this Killing spinor $(\epsilon^{(-)}_{\mf1})^*$ to find out the BPS relation of charges in Table \ref{chargeKS-}.%
\footnote{ The reason why we determine charges of $(\epsilon^{(-)}_{\mf1})^*$ instead of those of $\epsilon^{(-)}_{\mf1}$ is because 
$(\epsilon^{(-)}_{\mf1})^*$ has the negative energy, which has a corresponding supersymmetry charge, while $\epsilon^{(-)}_{\mf1}$ has the positive energy, which corresponds to a special superconformal charge. We fix a BPS relation of charges for a particular supersymmetry charge. }
\begin{table}[h]
\begin{center}
\begin{tabular}{cccccccccc}
\hline
  &  $H$ & $J_\phi$  & $J_\psi$ & $R^{\mf1}{}_{\mf1}{}$ & $R^{\mf2}{}_{\mf2}{}$& $R^{\mf3}{}_{\mf3}{}$  &$ R^{\mf4}{}_{\mf4}{}$ \\
\hline
\hline
$(\epsilon^{(-)}_{\mf1})^*$ & $-{1\over 2r} $ & ${w^{(-)} \over 2}$ & $0$& $-{3 \over 4}$ & ${1 \over 4}$ & ${1 \over 4}$ & ${1 \over 4}$  \\
\hline
\end{tabular}
\caption{Charges of the Killing spinor $(\epsilon^{(-)}_{\mf1})^*$ are shown. }
\label{chargeKS-}
\end{center}
\end{table}
Therefore the BPS relation of charges associated with the Killing spinor $\epsilon^{(-)}_{\mf1}$ is 
\be
r H = 2 R^{\mf1}{}_{\mf1} + 2  w^{(-)} J_\phi.
\label{bpscharge2}
\ee

To determine a BPS condition preserving $\epsilon^{(-)}_\mf1$, 
we extract the terms containing $\eta^{(-)}_\mf1$ from \eqref{dec1}, \eqref{dec2} as done previously. 
The results are as follows. 
\beal{\hspace{-2cm}
&\Delta_{{(-)}} \lambda_{A}\nn
=& e^{- {i\mu \over 4} t}e^{- iw^{(-)} \phi \over 2} \times \nn 
&\bigg[   \big\{(- iw^{(-)}c_\theta)  ( F^{01} - i F^{23})-   c_\theta  (F^{02} + i F^{13}) + s_\theta (F^{03} -i F^{12} ) -2 i[X_{AB}, X^{B\mf1}] ( - w^{(-)} s_\theta) \big\} \gamma_{1}\eta^{(-)}_{\mf1} \nn
&+\big\{ - i s_\theta  ( F^{01} - i F^{23}) + s_\theta w^{(-)}   (F^{02} + i F^{13})+c_\theta w^{(-)}  (F^{03} -i F^{12} )  -2 i[X_{AB}, X^{B\mf1}] c_\theta \big\} \eta^{(-)}_{\mf1} \bigg] +\cdots \nn
&\bar\Delta_{{(-)}} \lambda_{A} \nn
=&e^{ {i\mu \over 4} t} e^{ iw^{(-)} \phi \over 2} \times \nn
&\biggl[ \big\{ (- 2 D_0 X_{A\mf1} -X_{A\mf1}{2i \over r} ) (i w^{(-)} c_\theta )  - 2 D_1 X_{A\mf1} ( - w^{(-)} s_\theta ) - 2 D_2 X_{A\mf1}( i s_\theta ) - 2 D_3 X_{A\mf1} (i c_\theta )  \big\} (\eta^{(-)}_\mf1)^* \nn
&+ \big\{ (- 2 D_0 X_{A\mf1} -X_{A\mf1}{2i \over r}) (i s_\theta ) - 2 D_1 X_{A\mf1}( c_\theta  )   - 2 D_2 X_{A\mf1}( iw^{(-)} c_\theta ) - 2 D_3 X_{A\mf1} ( -i w^{(-)} s_\theta)\big\} \gamma_1 (\eta^{(-)}_\mf1)^* \bigg]\nn
&+\cdots \notag
}
where the ellipses describe the other terms than $\eta^{(-)}_\mf1$. 
In order for these to vanish with $\eta^{(-)}_{\mathfrak 1}$ nonzero, 
the fields are required to satisfy 
 \beal{
 &\big\{ - i s_\theta  ( F^{01} - i F^{23}) + s_\theta w^{(-)}   (F^{02} + i F^{13})+c_\theta w^{(-)}  (F^{03} -i F^{12} )\big\} \delta^{\mf1}_A   -2 i[X_{AB}, X^{B\mf1}] c_\theta=0,\nn
 &\big\{ (- iw^{(-)}c_\theta)  ( F^{01} - i F^{23})-   c_\theta  (F^{02} + i F^{13}) + s_\theta (F^{03} -i F^{12} )  \big\} \delta^{\mf1}_A -2 i[X_{AB}, X^{B\mf1}] (- w^{(-)} s_\theta)=0,\nn
& (- 2 D_0 X_{A\mf1} -X_{A\mf1}{2i \over r} ) (i w^{(-)} c_\theta )  - 2 D_1 X_{A\mf1} ( - w^{(-)} s_\theta ) - 2 D_2 X_{A\mf1}( i s_\theta ) - 2 D_3 X_{A\mf1} (i c_\theta ) =0, \nn
& (- 2 D_0 X_{A\mf1} -X_{A\mf1}{2i \over r}) (i s_\theta ) - 2 D_1 X_{A\mf1}( c_\theta  ) - 2 D_2 X_{A\mf1}( iw^{(-)} c_\theta ) - 2 D_3 X_{A\mf1} ( - i w^{(-)} s_\theta)=0,\notag
 }
for all $A = \mf1, \mf2, \mf3, \mf4$, since $\eta^{(-)}_{\mf1}, \gamma_1 \eta^{(-)}_{\mf1}, ( \eta^{(-)}_{\mf1})^*, \gamma_1 (\eta^{(-)}_{\mf1})^*$ are linearly independent under the projection \eqref{projection-}. 
Using the fact that all the components of $F^{\mu\nu}$ are real,  
we can simplify these BPS equations as follows. 
\begin{equation}\begin{split}
&F^{02}=- w^{(-)} F^{23} \sec \theta +  F^{03} \tan \theta,\quad
F^{13}= - w^{(-)} F^{01} \sec \theta  - F^{12} \tan \theta,\\
&[X_{AB}, X^{B\mf1}] = {1 \over 2i} \big\{ w^{(-)} \sec \theta  (F^{03}-i F^{12}) - (F^{23}+i F^{01}) \tan \theta \big\} \delta^{\mf1}_A,  \\
&D_0 X_{A\mf1}= - {i\over r} X_{A\mf1} - w^{(-)}  {2\over r} D_\phi X_{A\mf1},\\
&D_1 X_{A\mf1}= i w^{(-)} (-{ \cos\theta} D_2 X_{A\mf1}  + \sin\theta D_3 X_{A\mf1}).
\label{bps-general}
\end{split}\end{equation}
These are the other set of $1/16$ BPS equations, which is a new result in this paper. 

Let us solve these BPS equations when the fields take values in Cartan subalgebra of a gauge group.  
In this case these BPS equations are further simplified as
\beal{
&F^{02} = - w^{(-)}  \cos\theta F^{23}, \quad  
F^{03} =w^{(-)}   \sin\theta F^{23}, 
\label{bpsF-1}\\
&F^{12} = -w^{(-)}  \sin\theta F^{01}, \quad  
F^{13} = - w^{(-)}  \cos\theta F^{01},
\label{bpsF-2}\\
&(\partial_0 + {i\over r} + {2\over r} w^{(-)} \partial_\phi) X_{A\mf1}=0, \quad
\label{1/16BPS-D0X}\\
&(\partial_1 - i w^{(-)} (-{ \cos\theta} \partial_2  + \sin\theta \partial_3)) X_{A\mf1}=0.
\label{1/16BPS-D1X}
}
Let us first solve the matter BPS equation. 
By noticing that \eqref{1/16BPS-D1X} can be written as 
$
\hat L_{w^{(-)}} X_{A\mf1} =0,
$
where $\hat L_{w^{(-)}}=\hat L_{\pm}$ when $w^{(-)}=\pm1$, 
the BPS solution of the matter fields are expanded by the spherical harmonics of highest (or lowest) weight of $su(2)_L$ for $w^{(-)}=+1$ (or $-1$).
\be
X_{A\mf1} = \sum_{s\geq0 , |r_3|\leq s} x_{A\mf1}^{s, r_3}(t) Y_{s, w^{(-)} s, r_3}(\theta,\phi,\psi)
\ee
where $x_{A\mf1}^{s, r_3}(t)$ represents time-dependence of the scalar field, 
which is determined from \eqref{1/16BPS-D0X}. 
\be
\partial_0 x_{A\mf1}^{s, r_3}(t)=  {- i\over r}x_{A\mf1}^{s, r_3}(t) - {2 \over r} w^{(-)} (i w^{(-)} s) x_{A\mf1}^{s, r_3}(t).
\ee
This equation is easily solved as 
\be
x_{A\mf1}^{s, r_3}(t) = x_{A\mf1}^{s, r_3} e^{-i {2s+1\over r}t}
\ee
where $x_{A\mf1}^{s, r_3}$ is an integral constant. 
As a result we obtain BPS solutions 
\be
X_{A\mf1} = \sum_{s\geq0 , |r_3|\leq s} x_{A\mf1}^{s, r_3} e^{-i {2s+1\over r}t}Y_{s, w^{(-)} s, r_3}(\theta,\phi,\psi). 
\label{1/16BPS-Xsol}
\ee
This is a general $1/16$ BPS solution which preserves $\epsilon^{(-)}_\mf1$ projected by \eqref{projection+}. 
It is not difficult to check that this satisfies the equation of motion of the scalar field \eqref{eomX}. 

Let us move on to determining the BPS solution of the gauge sector
combining the equations of motion \eqref{eomF} and Bianchi identity. 
In the current situation they are given by  
$
\nabla_{\mu}F^{\mu\nu}=0
$
and 
$
\nabla_1 F^{23}+\nabla_2 F^{31}+\nabla_3 F^{12} =0, 
$
respectively. 
Under the BPS conditions \eqref{bpsF-1},\eqref{bpsF-2} 
the equations of motion reduce to the following three equations  
\beal{
\partial_{t} F^{01} =&- w^{(-)} {2 \over r} \partial_{\phi}  F^{01},
\label{1/16BPS-D0F01}\\
\partial_t F^{23} =& - w^{(-)} {2 \over r} \partial_\phi F^{23}, 
\label{1/16BPS-D0F23}\\
\partial_{3} F^{23} =& - w^{(-)}( \partial_{t} (\cos\theta F^{23}) +\partial_{1}(\sin\theta F^{01}) ),
\label{1/16BPS-D3F23}
}
and the Bianchi identity becomes 
\be
(-{1\over\sin\theta} \partial_{\psi} +\cot\theta\partial_\phi ) (\sin\theta F^{01}) =- w^{(-)}\partial_{\theta} (\sin\theta  F^{23}).
\label{bianchibps2}
\ee
We can solve \eqref{1/16BPS-D3F23} and \eqref{bianchibps2} 
by noticing the fact that they can be rewritten as
$\hat L_{w^{(-)}} G=0$, where 
$G = \sin\theta F^{01} + i \sin\theta F^{23}$.
Therefore this solution is given by the spherical harmonics of the highest (or lowest) weight 
of $su(2)_L$ algebra. 
\be
G =   \sum_{s\geq\half , |l_3|\leq s} C^{(-)}_{s, r_3}(t) Y_{s, w^{(-)} s, r_3}(\theta,\phi,\psi)
\ee
where $C^{(-)}_{s, r_3}(t)$ is an unknown function of time, 
which can be easily determined from the other BPS equations \eqref{1/16BPS-D0F01}, 
\eqref{1/16BPS-D0F23}.
\beal{
\partial_t C^{(-)}_{s, r_3}(t)  =- i {2 \over r} s C^{(-)}_{s, r_3}(t) 
}
which is solved as 
\be
C^{(-)}_{s, r_3} (t)= C^{(-)}_{s, r_3} e^{-i {2 \over r} s t} 
= a^{(-)}_{s, r_3} e^{-i {2 \over r} s t+ i\alpha^{(-)}_{s, r_3} } 
\ee
where $a^{(-)}_{s, r_3}, \alpha^{(-)}_{s, r_3}$ are integral constants with the range 
$a^{(-)}_{s, r_3} \geq 0, ~ 0 \leq \alpha^{(-)}_{s, r_3} < 2\pi$. 
Plugging this back into the above gives
\be
G= \sum_{s\geq\half, |l_3|\leq s}  a^{(-)}_{s, r_3} e^{- i {2 \over r} st+ i\alpha^{(-)}_{s, r_3} }  Y_{s, w^{(-)} s, r_3}(\theta,\phi,\psi).
\ee
Plugging the polar coordinates expression of the spherical harmonics given by \eqref{su2Lhighest} 
we find the $1/16$ BPS solutions of the field strength 
\beal{\hspace{-.5cm}
F^{01}
= & \sum_{s\geq\half, |r_3|\leq s} a^{(-)}_{s, r_3} c_{s, r_3} \tan^{w^{(-)} r_3}\left(\frac{\theta }{2}\right) \sin ^{s-1}\theta \cos{(-r_3\psi + s(w^{(-)} \phi- {2 \over r} t)+ \alpha^{(-)}_{s, r_3})}, 
\label{1/16BPS-F01sol}\\
\hspace{-.5cm}
F^{23}= & \sum_{s\geq\half, |r_3|\leq s} a^{(-)}_{s, r_3} c_{s, r_3} \tan^{w^{(-)} r_3}\left(\frac{\theta }{2}\right) \sin ^{s-1}\theta \sin(-r_3\psi + s(w^{(-)} \phi- {2 \over r} t)+ \alpha^{(-)}_{s, r_3}), 
\label{1/16BPS-F23sol}
}
where $c_{s, r_3}$ is given by \eqref{csr3}.%
\footnote{
We exclude the mode with $s=0$.   
} 
Other components can be obtained from \eqref{bpsF-1}, \eqref{bpsF-2}.

The moduli space of the $1/16$ BPS solution is given by
\be
x_{\mf1\mf2}^{s, r_3},x_{\mf1\mf2}^{s, r_3},x_{\mf1\mf3}^{s, r_3} \in \mbf C, 
~ a^{(-)}_{s, r_3} \geq 0, ~ 0\leq \alpha^{(-)}_{s, r_3}< 2\pi
\label{modulispace-}
\ee 
for $s\geq0, |r_3|\leq s$ for $x_{A\mf4}^{s, r_3}$, $s\geq\half, |r_3|\leq s$ for $a^{(+)}_{s, r_3}$.
Here we do not take into account the flux quantization condition. 

Let us compute conserved charges \eqref{energy}, \eqref{momenta}, \eqref{Rcharge} of the BPS solution.
For this end we first simplify them by using the BPS conditions \eqref{bpsF-1}, \eqref{bpsF-2}, \eqref{1/16BPS-D0X}, \eqref{1/16BPS-D1X}.
The results of the matter part are the following. 
\beal{
H|_{F_{\mu\nu}=0}
=& {4\over g^2} \int d^3\Omega~  \Tr\bigg( {1\over r^2} |X_{A\mf1}|^2 +  | \partial_2 X_{A\mf1}|^2 + | \partial_3 X_{A\mf1}|^2+ { 2 i\over r^2}w^{(-)} X_{A\mf1} \partial_\phi X^{A\mf1}  \bigg), \\
J_{\theta}|_{F_{\mu\nu}=0} =& 0, \\
J_{\phi}|_{F_{\mu\nu}=0} =& \int d^3\Omega~   {1\over g^2}  \Tr \( -2  ( - {i\over r} X_{A\mf1} - w^{(-)} {2 \over r} D_\phi X_{A\mf1})) D_{\phi} X^{A\mf1} +(c.c.) \),  \nn
J_{\psi}|_{F_{\mu\nu}=0} =&\int d^3\Omega~   {1\over g^2} \Tr \(-2 ( - {i\over r} X_{A\mf1} -w^{(-)} {2\over r} D_\phi X_{A\mf1}) D_\psi X^{A\mf1} +(c.c.) \),  \\
R^{\mf1}{}_{\mf1}|_{F_{\mu\nu}=0} 
 =& i  \int d^3\Omega~   {2\over g^2}\Tr \(-{ i\over r} |X_{\mf1A}|^2  - w^{(-)} {2 \over r} X^{\mf1A} D_\phi X_{A\mf1}  \). 
}
The gauge sector is as follows. 
\beal{
H|_{X_{AB}=0}  =&  \int d^3\Omega~  {1\over g^2} \Tr\((F^{01})^2+ (F^{23})^2 \), \\
J_{\theta}|_{X_{AB}=0} =& 0, \\
J_{\phi}|_{X_{AB}=0}=&  w^{(-)} {r \over 2 g^2}  \int d^3\Omega~ \Tr \((F^{01})^2  + ( F^{23})^2 \), \\
J_{\psi}|_{X_{AB}=0} =&w^{(-)}  {r \over 2g^2} \int d^3\Omega~  \cos\theta  \Tr \((F^{01})^2  + ( F^{23})^2 \),  
}
with $R^{\mf1}{}_{\mf1}|_{X_{AB}=0} =0$. 
Note that at this stage we see the BPS relation of charges given by \eqref{bpscharge2} in the gauge sector. 

Let us compute these conserved charges under the BPS solution \eqref{1/16BPS-Xsol}. 
\eqref{1/16BPS-F01sol}, \eqref{1/16BPS-F23sol}. 
The matter parts are%
\footnote{ 
We note useful formulas 
\beal{
 \int d^3\Omega~   {1\over\sin^2\theta} |Y_{s, w^{(-)} s, r_3}|^2=& \frac{s (1+2 s)}{2 \left(-r_3^2+ s^2\right)}, \\
 \int d^3\Omega~    {\cot\theta\over\sin\theta}  |Y_{s, w^{(-)} s, r_3}|^2=&\frac{w^{(-)}r_3(1+2 s)}{2 (r_3^2- s^2)},\\
 \int d^3\Omega~  \cot^2\theta |Y_{s, w^{(-)} s, r_3}|^2=&-\frac{2 r_3^2+ s}{2( r_3^2-s^2)}, 
}
where $|r_3|<s$. We however used these formulas formally at $|r_3|=s$ in the main text. 
}
\beal{
H|_{F_{\mu\nu}=0}
=&{8 \over r^2g^2}\sum_{s\geq0 , |r_3|\leq s}  (2s+1)^2  \Tr|x_{A\mf1}^{s, r_3}|^2, \\
J_{\phi} |_{F_{\mu\nu}=0}=& {4 \over r g^2}  w^{(-)} \sum_{s\geq0 , |r_3|\leq s} (2s +1 ) s \Tr |x_{A\mf1}^{s, r_3}|^2,  \\
J_{\psi} |_{F_{\mu\nu}=0}=& {4 \over r g^2} \sum_{s\geq0 , |r_3|\leq s} - (2s+1 ) r_3 \Tr |x_{A\mf1}^{s, r_3}|^2, \\  
R^{\mf1}{}_{\mf1}|_{F_{\mu\nu}=0} =& {2\over rg^2}  \sum_{s\geq0 , |r_3|\leq s} (2s +1 ) \Tr|x_{A\mf1}^{s, r_3} |^2, 
}
which satisfies the BPS relation of charges given by \eqref{bpscharge2}. 
The gauge field part is the following. 
\beal{
&H|_{X_{AB}=0} =w^{(-)} {2 \over r} J_{\phi}|_{X_{AB}=0}
= {1\over g^2}\sum_{s\geq\half, |r_3|\leq s}   \frac{s (1+2 s)}{2 \left(s^2-r_3^2\right)} \Tr  (a^{(-)}_{s, r_3})^2, \\
& J_{\psi}|_{X_{AB}=0}
= {r \over 2g^2}  \sum_{s\geq\half, |r_3|\leq s} \frac{ r_3(1+2 s)}{2 (r_3^2- s^2)}\Tr (a^{(-)}_{s, r_3})^2.
}
Note that the modes with $|r_3|=s$ give divergent contribution to the energy and momenta. 
 
\subsection{Counting of supersymmetries } 
\label{countSUSY} 

In this subsection we count number of supersymmetries preserved by BPS solutions constructed in the
previous subsections. 
Let us count the number of supersymmetries of the BPS solutions given by \eqref{1/16BPS+Xsol}, \eqref{1/16BPS+F01sol}, \eqref{1/16BPS+F23sol}, 
which preserves at least $\eta^{(+)}_{\mf4}$ projected by \eqref{projection+}. 
For convenience we write them down here again. 
\beal{
X_{A\mf4} =& \sum_{s\geq0 , |l_3|\leq s} x_{A\mf4}^{s, l_3} e^{i {2s+1\over r}t}Y_{s, l_3, w^{(+)} s}(\theta,\phi,\psi),
\label{xa4}\\
F^{01}
= &  \sum_{s\geq\half , |l_3|\leq s} a^{(+)}_{s, l_3} c_{s,l_3} \tan^{w^{(+)} l_3}\left(\frac{\theta }{2}\right) \sin ^{s-1}\theta \cos{(l_3\phi  +s({2 \over r} t-w^{(+)}  \psi)+\alpha^{(+)}_{s, l_3})}, 
\label{f01mf4}\\
F^{23}= & \sum_{s\geq\half , |l_3|\leq s} a^{(+)}_{s, l_3} c_{s,l_3} \tan^{w^{(+)} l_3}\left(\frac{\theta }{2}\right) \sin ^{s-1}\theta \sin{(l_3\phi +s({2 \over r} t-w^{(+)}  \psi)+\alpha^{(+)}_{s, l_3})}, 
\label{f23mf4}
}
for arbitrary $A=\mf1, \mf2, \mf3$. 
To exclude the case of trivial angular momenta, 
we consider the case whether $x_{C\mf4}^{s, l_3} \not=0$ or $a^{(+)}_{s, l_3}\not=0$ for some positive $s$ and $ l_3 $. 

First let us consider the case where $a^{(+)}_{s, l_3}\not=0$ for some positive $s$ and $ l_3 $. 
Under this situation we consider the BPS solution which also preserves $\eta^{(+)}_{\mf3}$ satisfying $\gamma_0\eta^{(+)}_{\mf3}=i w^{(+)}_{\mf3} \eta^{(+)}_{\mf3}$.
Then the form of the BPS solution gets another constraint such that  
\beal{
X_{A\mf3} =& \sum_{s\geq0 , |l_3|\leq s} x_{A\mf3}^{s, l_3} e^{i {2s+1\over r}t}Y_{s, l_3, w^{(+)}_{\mf3} s}(\theta,\phi,\psi),
\label{xa3}\\
F^{01}
= &  \sum_{s\geq\half , |l_3|\leq s} a^{(+)}_{s, l_3} c_{s,l_3} \tan^{w^{(+)}_{\mf3} l_3}\left(\frac{\theta }{2}\right) \sin ^{s-1}\theta \cos{(l_3\phi  +s({2 \over r} t - w^{(+)}_{\mf3}  \psi)+\alpha^{(+)}_{s, l_3})},
\label{f01mf3}\\
F^{23}= & \sum_{s\geq\half , |l_3|\leq s} a^{(+)}_{s, l_3} c_{s,l_3} \tan^{w^{(+)}_{\mf3} l_3}\left(\frac{\theta }{2}\right) \sin ^{s-1}\theta \sin{(l_3\phi +s({2 \over r} t-w^{(+)}_{\mf3}  \psi)+\alpha^{(+)}_{s, l_3})},
\label{f23mf3}
}
for $A=\mf1, \mf2, \mf4$. 
Under the assumption, matching of \eqref{f01mf4}, \eqref{f23mf4} with \eqref{f01mf3}, \eqref{f23mf3} 
requires the signature of the projection for $\eta^{(+)}_{\mf3}$, $\eta^{(+)}_{\mf4}$ to be the same, 
$w^{(+)}_{\mf3} = w^{(+)}$. Matching of the matter solution \eqref{xa4}, \eqref{xa3} demands  
$X_{\mf1\mf4} = X_{\mf2\mf4}=0$ since 
$X_{\mf1\mf4}^\dagger=-X_{\mf2\mf3}, X_{\mf2\mf4}^\dagger=X_{\mf1\mf3}$ from \eqref{Xconstraint}.  
Under this situation supersymmetries specified by  $\eta^{(+)}_{\mf1}$ and $\eta^{(+)}_{\mf2}$ are broken unless $ X_{\mf3\mf4}^\dagger=0$. In other words, the matter becomes trivial if we keep three supersymmetries of the form $\eta^{(+)}_A$. Can the BPS solution preserve $\eta^{(-)}_A$ supersymmetry concurrently? 
Let us study this possibility. 
Keeping also $\eta^{(-)}_\mf4$ makes the matter fields completely trivial 
and gives constraint for the field strength
so that the highest or lowest weight modes in terms of $su(2)_L$ remain: 
$a^{(+)}_{s, l_3}=0$ for $\forall s>0, l_3 \not= - w^{(-)}s$ for fixed $ w^{(-)}$.
To preserve $\eta^{(-)}_\mf1$ in stead of $\eta^{(-)}_\mf4$, the matter field $X_{\mf1\mf4}$ has to vanish
and the other matter fields have only the modes which are also the highest or lowest weight states of $su(2)_L$.
In this case projected $\eta^{(-)}_\mf1$ is preserved. The condition to keep $\eta^{(-)}_\mf2$, $\eta^{(-)}_\mf3$ are the same with that of $\eta^{(-)}_\mf1$ by replacing $\mf1$ by $\mf2,\mf3$ respectively. 
When all scalar fields become trivial, all Killing spinors $\eta^{(+)}_A$ are preserved for all $A$. 
In this case, supersymmetry is enhanced when the gauge field strength is expanded by 
the mode of the highest or lowest weight representation of $su(2)_L$ as well 
so that $a^{(+)}_{s, l_3}=0$ for $\forall s>0, l_3 \not= - w^{(-)}s$ for fixed $ w^{(-)}$. 
In this case the Killing spinors of the form $\eta^{(-)}_A$ are also conserved. 
We summarize this result in Table \ref{table+1}. 
\begin{table}[htbp]
\begin{center}
\begin{tabular}{c|c|c}
\hline
Parameter region. &  Preserved Killing spinors. & Number of SUSY.  \\ \hline \hline
$x_{\mf1\mf4}^{s, l_3}, x_{\mf2\mf4}^{s, l_3}, x_{\mf3\mf4}^{s, l_3},a^{(+)}_{s, l_3}$: generic.   &$\eta^{(+)}_{\mf4}$   &  $2$  \\
   &with  $\gamma_0 \eta^{(+)}_{\mf4} = i w^{(+)}\eta^{(+)}_{\mf4}$. & (${1 \over 16}$ BPS) \\
\hline
$x_{\mf1\mf4}^{s, l_3}=x_{\mf2\mf4}^{s, l_3}=0$  for $\forall s, l_3$.  & $\eta^{(+)}_{\mf3},\eta^{(+)}_{\mf4}$   &  $4$  \\
 &with  $\gamma_0 \eta^{(+)}_{A} = i w^{(+)}\eta^{(+)}_{A}$. & (${1 \over 8}$ BPS) \\
\hline
 $x_{\mf2\mf4}^{s, l_3}, x_{\mf3\mf4}^{s, l_3}, a^{(+)}_{s, l_3} =0$ for $\forall s>0, l_3 \not= - w^{(-)}s$, & $\eta^{(-)}_{\mf1},\eta^{(+)}_{\mf4}$  &  $4$  \\
  $x_{\mf1\mf4}^{s, l_3}=0$  for $\forall s, l_3$.    &with  $\gamma_0 \eta^{(\pm)}_{A} = i w^{(\pm)}\eta^{(\pm)}_{A}$. & (${1 \over 8}$ BPS) \\
\hline
$x_{\mf3\mf4}^{s, l_3},a^{(+)}_{s, l_3}=0$ for $\forall s>0, l_3 \not= - w^{(-)}s$, & $\eta^{(-)}_{\mf1},\eta^{(-)}_{\mf2},\eta^{(+)}_{\mf3},\eta^{(+)}_{\mf4}$ &  $8$  \\
$x_{\mf1\mf4}^{s, l_3}=x_{\mf2\mf4}^{s, l_3}=0$ for $\forall s, l_3$. &with  $\gamma_0 \eta^{(\pm)}_{A} = i w^{(\pm)}\eta^{(\pm)}_{A}$. & (${1 \over 4}$ BPS) \\
\hline
$x_{\mf1\mf4}^{s, l_3}=x_{\mf2\mf4}^{s, l_3}=x_{\mf3\mf4}^{s, l_3}=0$ for $\forall s, l_3$. & $\eta^{(+)}_{\mf1},\eta^{(+)}_{\mf2},\eta^{(+)}_{\mf3},\eta^{(+)}_{\mf4}$ &  $8$  \\
 &with  $\gamma_0 \eta^{(+)}_{A} = i w^{(+)}\eta^{(+)}_{A}$. & (${1 \over 4}$ BPS) \\
\hline
$a^{(+)}_{s, l_3}=0$ for $\forall s>0, l_3 \not= - w^{(-)}s$, & $\eta^{(\pm)}_{\mf1},\eta^{(\pm)}_{\mf2},\eta^{(\pm)}_{\mf3},\eta^{(\pm)}_{\mf4}$ &  $16$  \\
$x_{\mf1\mf4}^{s, l_3}=x_{\mf2\mf4}^{s, l_3}=x_{\mf3\mf4}^{s, l_3}=0$.   for $\forall s, l_3$&  with  $\gamma_0 \eta^{(\pm)}_{A} = i w^{(\pm)}\eta^{(\pm)}_{A}$. & (${1 \over 2}$ BPS) \\
\hline
\end{tabular}
\caption{ We list the number of supersymmetry of BPS solutions \eqref{1/16BPS+Xsol}, \eqref{1/16BPS+F01sol}, \eqref{1/16BPS+F23sol} to preserve $\eta^{(+)}_{\mf4}$ with $\gamma_0 \eta^{(+)}_{\mf4} = i w^{(+)}\eta^{(+)}_{\mf4}$ at each point of moduli space with $a^{(+)}_{s, l_3}\not=0$ for $\exists s>0, l_3$. 
}
\label{table+1}
\end{center}
\end{table}

Next we consider a case where $x_{\mf3\mf4}^{s, l_3} \not=0$ for some positive $s$ and $ l_3 $. 
This assumption restricts us to two case to study BPS solutions to preserve another Killing spinor as 
(i) $\eta^{(+)}_{\mf3}$,  
(ii) $\eta^{(-)}_{\mf1}$. 
They are all projected by $\gamma_0$.
Note that the assumption excludes the cases $\eta^{(+)}_{\mf2}$ (or  $\eta^{(+)}_{\mf1}$), $\eta^{(-)}_{\mf3}$, $\eta^{(-)}_{\mf4}$, which result in $X_{\mf3\mf4}=0$.
(i) Let us first study BPS solutions to preserves another supersymmetry specified by 
$\eta^{(+)}_{\mf3}$ which satisfies $\gamma_0\eta^{(+)}_{\mf3}=i w^{(+)}_{\mf3} \eta^{(+)}_{\mf3}$. 
We study the BPS solution of the matter fields
since the constraint of the field strength can be discussed in the same way as above.
As in the above case, the matter BPS solution should also be of the form such as \eqref{xa3}. 
Due to the assumption, matching of \eqref{xa4} and \eqref{xa3} requires the signature of the projection for $\eta^{(+)}_{\mf3}$, $\eta^{(+)}_{\mf4}$ to be the same, 
$w^{(+)}_{\mf3} = w^{(+)}$, and $X_{\mf1\mf4} = X_{\mf2\mf4}=0$ from \eqref{Xconstraint}.  
It is not possible to preserve $\eta^{(+)}_\mf1$ or $\eta^{(+)}_\mf2$, which would make all matter fields trivial, though it is possible to conserve  $\eta^{(-)}_\mf1$ projected by $\gamma_0$, 
in which case the non-trivial scalar field $X_{\mf3\mf4}$ 
and the field strength have to also be expanded by the highest weight states of $su(2)_L$:
 $x_{\mf3\mf4}^{s, l_3}, a^{(+)}_{s, l_3} =0$ for $l_3 \not= - w^{(-)}s$. 
 This solution automatically has supersymmetry specified by $\eta^{(-)}_\mf2$ as well. 
(ii) Let us consider the case for the BPS solution to preserves 
$\eta^{(-)}_{\mf1}$ which satisfies $\gamma_0\eta^{(-)}_{\mf1}=i w^{(-)} \eta^{(-)}_{\mf1}$.
This demands that $X_{\mf1\mf4}=0$ and $x_{\mf2\mf4}^{s, l_3},x_{\mf3\mf4}^{s, l_3}, a^{(+)}_{s, l_3} =0$ for $l_3 \not= - w^{(-)}s$. 
Notice that supersymmetry is enhanced when  $x_{\mf2\mf4}^{s, l_3},a^{(+)}_{s, l_3} =0$ for all $s>0$, in which case $\eta^{(-)}_{\mf1}$ is preserved without any projection. 
Furthermore in order to preserve $\eta^{(-)}_{\mf2}$ in addition it is necessary for $X_{\mf2\mf4}$ to vanish.
In this case supersymmetry is enhanced so that $\eta^{(+)}_{\mf3}$ is also preserved.
We summarize this result in Table \ref{table+2}. 
\begin{table}[htbp]
\begin{center}
\begin{tabular}{c|c|c}
\hline
Parameter region. &  Preserved Killing spinors. & Number of SUSY.  \\ \hline \hline
$x_{\mf1\mf4}^{s, l_3}, x_{\mf2\mf4}^{s, l_3}, x_{\mf3\mf4}^{s, l_3},a^{(+)}_{s, l_3}$: generic.   &$\eta^{(+)}_{\mf4}$ &  $2$  \\
   &with  $\gamma_0 \eta^{(+)}_{\mf4} = i w^{(+)}\eta^{(+)}_{\mf4}$. & (${1 \over 16}$ BPS) \\
\hline
  $x_{\mf1\mf4}^{s, l_3}=x_{\mf2\mf4}^{s, l_3}=0$  for $\forall s, l_3$.   & $\eta^{(+)}_{\mf3},\eta^{(+)}_{\mf4}$  &  $4$  \\
&with  $\gamma_0 \eta^{(+)}_{A} = i w^{(+)}\eta^{(+)}_{A}$.  & (${1 \over 8}$ BPS) \\
\hline
 $x_{\mf2\mf4}^{s, l_3}, x_{\mf3\mf4}^{s, l_3}, a^{(+)}_{s, l_3} =0$ for $\forall s>0, l_3 \not= - w^{(-)}s$,  & $\eta^{(-)}_{\mf1},\eta^{(+)}_{\mf4}$   &  $4$  \\
  $x_{\mf1\mf4}^{s, l_3}=0$  for $\forall s, l_3$.    &with  $\gamma_0 \eta^{(\pm)}_{A} = i w^{(\pm)}\eta^{(\pm)}_{A}$.  & (${1 \over 8}$ BPS) \\
\hline
 $x_{\mf3\mf4}^{s, l_3}=0$ for $\forall s>0, l_3 \not= - w^{(-)}s$  & $\eta^{(-)}_{\mf1},\eta^{(+)}_{\mf4}$  &  $6$  \\
  $x_{\mf2\mf4}^{s, l_3}, a^{(+)}_{s, l_3} =0$  for $\forall s>0, l_3$; $ x_{\mf1\mf4}^{s, l_3} =0$  for $\forall s, l_3$.    &with  $\gamma_0 \eta^{(+)}_{\mf4} = i w^{(+)}\eta^{(+)}_{\mf4}$. & (${3 \over 16}$ BPS) \\
\hline
 $x_{\mf3\mf4}^{s, l_3}, a^{(+)}_{s, l_3} =0$ for $\forall s>0, l_3 \not= - w^{(-)}s$,  & $\eta^{(-)}_{\mf1},\eta^{(-)}_{\mf2},\eta^{(+)}_{\mf3},\eta^{(+)}_{\mf4}$   &  $8$  \\
$x_{\mf1\mf4}^{s, l_3}=x_{\mf2\mf4}^{s, l_3}=0$  for $\forall s, l_3$.  &with  $\gamma_0 \eta^{(\pm)}_{A} = i w^{(\pm)}\eta^{(\pm)}_{A}$. & (${1 \over 4}$ BPS) \\
\hline
\end{tabular}
\caption{ We list the number of supersymmetries of BPS solutions \eqref{1/16BPS+Xsol}, \eqref{1/16BPS+F01sol}, \eqref{1/16BPS+F23sol} to preserve $\eta^{(+)}_{\mf4}$ with $\gamma_0 \eta^{(+)}_{\mf4} = i w^{(+)}\eta^{(+)}_{\mf4}$ at each point of moduli space with $x_{\mf3\mf4}^{s, l_3} \not=0$ for $\exists s>0, l_3 $.
}
\label{table+2}
\end{center}
\end{table}

Similarly we can count the number of supersymmetries of the BPS solution which preserves at least $\eta^{(-)}_\mf1$ with projection \eqref{projection-}. 
They are given by \eqref{1/16BPS-Xsol}, \eqref{1/16BPS-F01sol}, \eqref{1/16BPS-F23sol}, 
which we write down again for convenience. 
\beal{\hspace{-.5cm}
X_{A\mf1} =& \sum_{s\geq0 , |r_3|\leq s} x_{A\mf1}^{s, r_3} e^{-i {2s+1\over r}t}Y_{s, w^{(-)}_{\mf3} s, r_3}(\theta,\phi,\psi), \\
\hspace{-.5cm}
F^{01}
= & \sum_{s\geq\half, |r_3|\leq s} a^{(-)}_{s, r_3} c_{s, r_3} \tan^{w^{(-)} r_3}\left(\frac{\theta }{2}\right) \sin ^{s-1}\theta \cos{(-r_3\psi + s(w^{(-)} \phi- {2 \over r} t)+ \alpha^{(-)}_{s, r_3})}, \\
\hspace{-.5cm}
F^{23}= & \sum_{s\geq\half, |r_3|\leq s} a^{(-)}_{s, r_3} c_{s, r_3} \tan^{w^{(-)} r_3}\left(\frac{\theta }{2}\right) \sin ^{s-1}\theta \sin(-r_3\psi + s(w^{(-)} \phi- {2 \over r} t)+ \alpha^{(-)}_{s, r_3}). 
}
Discussion how to count the number of supersymmetries can be done in a parallel way, 
so we only present results of the table corresponding to 
Table \ref{sphericalBPS}, Table \ref{table+1}, Table \ref{table+2} by Table \ref{table-}. 

\begin{table}[htbp]
\begin{center}
\begin{tabular}{c|c|c}
\hline
Parameter region. &  Preserved Killing spinors. & Number of SUSY.  \\ \hline \hline
$x_{\mf1\mf2}^{s, r_3}, x_{\mf1\mf3}^{s, r_3}, x_{\mf1\mf4}^{s, r_3},a^{(-)}_{s, r_3}$: generic.   &$\eta^{(-) }_{\mf1}$   &  $2$  \\
   &with  $\gamma_0 \eta^{(-) }_{\mf1} = i w^{(-)}\eta^{(-) }_{\mf1}.$ & (${1 \over 16}$ BPS) \\
\hline
$x_{\mf1\mf3}^{s, r_3}=x_{\mf1\mf4}^{s, r_3}=0$  for $\forall s, r_3$.  & $\eta^{(-) }_{\mf1},\eta^{(-)}_{\mf2}$   &  $4$  \\
 &with  $\gamma_0 \eta^{(-)}_{A} = i w^{(-)}\eta^{(-)}_{A}.$ & (${1 \over 8}$ BPS) \\
\hline
 $x_{\mf1\mf2}^{s, r_3}, x_{\mf1\mf3}^{s, r_3}, a^{(-)}_{s, r_3} =0$ for $\forall s, r_3 \not= - w^{(+)}s$, & $\eta^{(-) }_{\mf1},\eta^{(+) }_{\mf4}$   &  $4$  \\
  $x_{\mf1\mf4}^{s, r_3}=0$  for $\forall s, r_3$. &with  $\gamma_0 \eta^{(\pm)}_{A} = i w^{(\pm)}\eta^{(\pm)}_{A}$. & (${1 \over 8}$ BPS) \\
\hline
 $x_{\mf1\mf2}^{s, r_3}, x_{\mf1\mf3}^{s, r_3}, x_{\mf1\mf4}^{s, r_3} =0$ for $\forall s>0, r_3$, & $\eta^{(-) }_{\mf1}.$   &  $4$  \\
  $a^{(-)}_{s, r_3}=0$  for $\forall s, r_3$.    &  & (${1 \over 8}$ BPS) \\
\hline
 $x_{\mf1\mf2}^{s, r_3}=0$ for $\forall s, r_3 \not= - w^{(+)}s$ & $\eta^{(-) }_{\mf1},\eta^{(+) }_{\mf4}$,   &  $6$  \\
  $x_{\mf1\mf3}^{s, r_3}, a^{(-)}_{s, r_3} =0$  for $\forall s>0, l_3$;   $ x_{\mf1\mf4}^{s, r_3}=0$  for $\forall s, l_3$.    &with  $\gamma_0 \eta^{(-) }_{\mf1} = i w^{(-)}\eta^{(-) }_{\mf1}$. & (${3 \over 16}$ BPS) \\
\hline
 $x_{\mf1\mf2}^{s, r_3},a^{(-)}_{s, r_3}=0$ for $\forall s, r_3 \not= - w^{(+)}s$, & $\eta^{(-) }_{\mf1},\eta^{(-)}_{\mf2},\eta^{(+)}_{\mf3},\eta^{(+) }_{\mf4}$   &  $8$  \\
$x_{\mf1\mf3}^{s, r_3}=x_{\mf1\mf4}^{s, r_3}=0$ for $\forall s, r_3$. &with  $\gamma_0 \eta^{(\pm)}_{A} = i w^{(\pm)}\eta^{(\pm)}_{A}$. & (${1 \over 4}$ BPS) \\
\hline
$x_{\mf1\mf2}^{s, r_3}=x_{\mf1\mf3}^{s, r_3}=x_{\mf1\mf4}^{s, r_3}=0$ for $\forall s, r_3$. & $\eta^{(-)}_{\mf1},\eta^{(-)}_{\mf2},\eta^{(-)}_{\mf3},\eta^{(-) }_{\mf4}$   &  $8$  \\
 &with  $\gamma_0 \eta^{(-)}_{A} = i w^{(-)}\eta^{(-)}_{A}$. & (${1 \over 4}$ BPS) \\
\hline
$x_{\mf1\mf2}^{s, r_3}=x_{\mf1\mf3}^{s, r_3}=0$ for $\forall s>0, r_3$, & $\eta^{(-)}_{\mf1}, \eta^{(+) }_{\mf4}$.   &  $8$  \\
 $x_{\mf1\mf4}^{s, r_3}=a^{(-)}_{s, r_3}=0$  for $\forall s, r_3$.  &  & (${1 \over 4}$ BPS) \\
\hline
$a^{(-)}_{s, r_3}=0$ for $\forall s, r_3 \not= - w^{(+)}s$, & $\eta^{(\pm)}_{\mf1},\eta^{(\pm)}_{\mf2},\eta^{(\pm)}_{\mf3},\eta^{(\pm)}_{\mf4}$   &  $16$  \\
$x_{\mf1\mf2}^{s, r_3}=x_{\mf1\mf3}^{s, r_3}=x_{\mf1\mf4}^{s, r_3}=0$ for $\forall s, r_3$.   &with  $\gamma_0 \eta^{(\pm)}_{A} = i w^{(\pm)}\eta^{(\pm)}_{A}$. & (${1 \over 2}$ BPS) \\
\hline
$x_{\mf1\mf2}^{s, r_3}=0$ for $\forall s>0, r_3$, &$\eta^{(-)}_{\mf1},\eta^{(-)}_{\mf2},\eta^{(+)}_{\mf3},\eta^{(+) }_{\mf4}$.  &  $16$  \\
 $x_{\mf1\mf3}^{s, r_3}=x_{\mf1\mf4}^{s, r_3}=a^{(-)}_{s, r_3}=0$  for $\forall s, r_3$.  &  & (${1 \over 2}$ BPS) \\
\hline
 $x_{\mf1\mf2}^{s, r_3}=x_{\mf1\mf3}^{s, r_3}=x_{\mf1\mf4}^{s, r_3}=a^{(-)}_{s, r_3}=0$  for $\forall s, r_3$.   &$\eta^{(\pm)}_{\mf1},\eta^{(\pm)}_{\mf2},\eta^{(\pm)}_{\mf3},\eta^{(\pm) }_{\mf4}$.  &  $32$  \\
&  & (Unique vacuum) \\
\hline
\end{tabular}
\caption{ We list the number of supersymmetry of BPS solutions \eqref{1/16BPS-Xsol}, \eqref{1/16BPS-F01sol}, \eqref{1/16BPS-F23sol} which preserve $\eta^{(-) }_{\mf1}$ with $\gamma_0 \eta^{(-) }_{\mf1} = i w^{(-)}\eta^{(-) }_{\mf1}$ at each point of moduli space.
}
\label{table-}
\end{center}
\end{table}

\section{BPS states in $\cN=8$ SYM on $\mbf R\times \mbf S^2$ } 
\label{BPSN=8SYM}

In this section we investigate supersymmetric states in $\cN=8$ SYM on $\mbf R\times \mbf S^2$
by carrying out dimensional reduction of $\cN=4$ SYM on $\mbf R\times \mbf S^3$
so that the Hopf fiber direction $\psi$ is degenerated.
We present basic results obtained by this dimensional reduction for this paper 
to be self-contained. 
From the metric of $\mbf R\times \mbf S^3$ we obtain that of $\mbf R\times \mbf S^2$ as 
\be
ds^2_{\mbf R\times \mbf S^2} =-dt^2 + \mu^{-2} ( d\theta^2+\sin^2\theta d\phi^2)
\ee
where we set $\mu={2\over r}$, which is the inverse radius of $\mbf S^2$. 
For more detail, see Appendix \ref{s2}.  
Dimensional reduction for fields can be achieved by truncating the fields 
to leave the zero modes of the Hopf fiber direction. 
$
\partial _\psi \Phi = 0, 
$
or
$
\partial _3 \Phi = 0, 
$
where $\Phi$ is any four dimensional field. 
Accordingly the four dimensional gauge field is separated into 
the three dimensional one and a scalar field in a way that
\bea
A_{i} = a_{i}, \quad A_{3}= \phi, 
\label{4dto3dgauge}
\eea
where $i=0,1,2$. 
Thus the gauge field strength become
\beal{
&F_{0 1} = f_{01}, \quad
F_{02} = f_{02}, \quad
F_{\mu3} = D_\mu \phi,\quad
F_{12} = f_{12}-\mu\phi, \quad
\label{4dto3dfieldstrength}
}
where $f_{\mu\nu}=\partial_\mu a_\nu-\partial_\nu a_\mu +i[a_\mu, a_\nu],$ 
$D_\mu \phi = \partial_\mu \phi+i[a_\mu,\phi]$ with $\mu, \nu=t, \theta, \phi$.
Note that the local Lorentz indices $0,1,2$ and the global indices $t, \theta, \phi$ 
are now transformed to each other by using the transition function of $\mbf R\times \mbf S^2$
given by \eqref{vielbeins2}.

Performing this dimensional reduction to the $\cN=4$ SYM action given by \eqref{N4SYMaction} 
we obtain the action of $\cN=8$ SYM on $\mbf R\times\mbf S^2$
\beal{
S=& {1\over g_3^2}\int dt {d^2\Omega\over \mu^2}  \Tr
\biggl[-{1\over 4} f_{\mu\nu} f^{\mu\nu}
-\half D_\mu \phi D^\mu \phi+\mu \phi f_{12}-\half \mu^2 \phi^2 \nn
&-\half D_\mu X_{AB} D^\mu X^{AB}
-{\mu^2\over 8} X_{AB}X^{AB}
+{1\over2}[X_{AB},\phi][X^{AB},\phi]
+{1\over4}[X_{AB},X_{CD}][X^{AB},X^{CD}]
\nn
&+{i}(\psi_A)^\dagger\gamma^\mu D_\mu \psi_A
+{\mu\over4}(\psi_A)^\dagger\rho^t \psi_A
+{i}(\psi_A)^\dagger [\phi, \psi_A] 
+(\psi_A)^\dagger [X_{AB}, (\psi_B)^\dagger]
+\psi_A [X^{AB},  \psi_{B}] 
\biggl]
\label{N=8SYMaction}
}
where we set $d^2\Omega= d\theta\sin\theta d\phi$, $\lambda_A=\psi_A$, 
$
{1\over g^2_{3}} = \frac{1}{g^2} {4 \pi \over \mu}, 
$
and $D_\mu$ is the covariant derivative in terms of gauge and spin indices in the three dimension. 

Equations of motion for the gauge field and the scalar fields in the bosonic part are 
\beal{
&D_\mu f^{\mu\nu} - i [\phi, D^\nu \phi] - i [X_{AB}, D^\nu X^{AB}] +\mu\gamma^{0\mu\nu}D_\mu\phi=0,
\label{eomf}\\
&D^\mu D_\mu \phi + \mu f_{12} -{\mu^2} \phi^2 +[X_{AB},[\phi,X^{AB}]]=0,
\label{eomphi}\\
&D^\mu D_\mu X^{AB} -{\mu^2\over 4} X^{AB}+[\phi,[X^{AB},\phi]]+[X_{CD},[X^{AB},X^{CD}]]=0.
}

Conserved charges of this theory can be obtained by dimensional reduction. 
$SU(4)_R$ R-symmetry charge of the bosonic part is 
\beal{
R^{C}{}_A =&i  \int d^2\Omega~   {1\over g_3^2} \Tr \( - X^{CB} D^0 X_{AB} + D^0 X^{CB} X_{AB} \). 
}
The energy is 
\beal{
H =&  \int d^2\Omega~  {1\over g_3^2} \Tr\bigg(\half (f^{0i})^2+ {1\over 4} (f^{ij})^2 +\half |D_0 \phi|^2 +\half |\vec D \phi|^2 +\half |D_0 X_{AB}|^2 +\half |\vec D X^{AB}|^2 \nn
&-\mu\phi f_{12} + \half \mu^2\phi^2 + {1\over 2} |[X_{AB},\phi]|^2 + {1\over 4} |[X_{AB},X_{CD}]|^2 +{\mu^2 \over8} |X_{AB}|^2 \bigg). 
}
The angular momentums are 
\beal{
P^{i} =& \int d^2\Omega~   {1\over g_3^2} \Tr \(f^{0}{}_\mu f^{i\mu} +D^{0}\phi D^{i}\phi  +\half(D^{0}X_{AB} D^{i} X^{AB} + D^{i}X_{AB} D^{0} X^{AB}) \) 
}
where $i = \theta, \phi$. 

Under the truncation supersymmetries specified by $\epsilon^{(+)}_A$ are all broken, 
since they are dependent on the Hopf fiber direction $\psi$.  
The other supersymmetries specified by $\epsilon^{(-)}_A$ are all preserved, 
which we denote by $\xi_A$. 
The determining equation of $\xi_A$ comes from \eqref{KSeq-}, which is now given by 
\be
\partial_{t}\xi_A = - {i \mu \over 4} \xi_A, \quad   
\nabla^{\mbf S^2}_{\theta}\xi_A = - {i \mu \over 2}\rho_{{\theta}}{}^t \xi_A, \quad 
\nabla^{\mbf S^2}_{\phi}\xi_A = - {i \mu \over 2}\rho_{{\phi}}{}^t \xi_A,  
\ee
and the solution is 
\be
\xi_{A} = e^{- i {\mu \over 4} t} e^{- { i \over 2} \rho_{01} \theta} e^{- \half \rho_{21} \phi} \eta_{A}
\ee
where $\eta_{A}$ is a constant spinor. 
By using this the supersymmetry transformation rule of $\cN=8$ SYM on $\mbf R\times \mbf S^2$ 
is obtained as
\beal{
\Delta_\xi A_{\mu} 
=&i ((\xi_A)^* \rho_{{\mu}} \psi_{A}
- (\psi_A)^\dagger \rho_{{\mu}} \xi_{A}), \nn
\Delta_\xi \phi
=& (\xi_A)^* \psi_{A} - (\psi_A)^\dagger \xi_{A}, \nn 
\Delta_\xi X^{AB} =& i( - \varepsilon^{ABCD} \xi_C \psi_{D}
- (\xi_A)^* (\psi_B)^\dagger+ (\xi_B)^* (\psi_A)^\dagger), \nn
\Delta_\xi\psi_A =& (-i D_\mu \phi \rho^\mu + \half f_{\mu\nu}\rho^{\mu\nu} - \mu\phi \rho^{12}) \xi_{A}
-2 ( D_\mu X_{AB} \rho^\mu - [\phi, X_{AB} ] ) (\xi_B)^* \nn
&- 2i[X_{AB}, X^{BC}] \xi_{C} - X_{AB}  (\rho^\mu \nabla^{\mbf S^2}_\mu - {i \mu \over 4} \rho^{12} )(\xi_B)^*.
\label{susyrs3}
}
Note that due to this reduction the global symmetry reduces from $PSU(2,2|4)$ to $PSU(2|4)$.

BPS states of $\cN=8$ SYM on $\mbf R\times \mbf S^2$ can be studied in the same manner 
as the case of $\cN=4$ SYM on $\mbf R\times \mbf S^3$ in \S\ref{1/16BPSstates-}. 
Especially calculation in \S\ref{1/16BPSstates-} can be applied to this case as it is 
by exchanging the fields from four dimension to three dimension as discussed above.  
For this reason we do not repeat the similar calculation, and we present only relevant results. 

Let us study BPS states of $\cN=8$ SYM on $\mbf R\times \mbf S^2$. 
For this purpose we study a BPS state which preserves a Killing spinor $\epsilon_\mf1$ whose constant spinor $\eta_{\mf1}$ is projected in such a way that 
\beal{
&\gamma_0 \eta_{\mf1}= i w\eta_{\mf1}
}
where $w= \pm 1$ and $\mf1=1,2,3,4$.
The BPS relation of charges associated with the Killing spinor $\epsilon_{\mf1}$ is 
\be
\mu^{-1} H = R^{\mf1}{}_{\mf1} +  w J_\phi.
\label{bpscharge3d}
\ee
By performing the dimensional reduction to \eqref{bps-general}, 
we find 
\begin{equation} \begin{split}
&f^{02}=- w D^{2} \phi \sec \theta +  D^{0}\phi \tan \theta, \quad 
D^{1} \phi = - w f^{01} \sec \theta  - ( f_{12}-\mu\phi) \tan \theta,\\
&[X_{AB}, X^{B\mf1}] = {1 \over 2i} \big\{ w \sec \theta  (D^{0}\phi -i ( f_{12}-\mu\phi) ) - (D^{2}\phi +i f^{01}) \tan \theta \big\} \delta^{\mf1}_A,  \\
&D_0 X_{A\mf1}= - {i\mu \over 2} X_{A\mf1} - w \mu D_\phi X_{A\mf1},\\
&D_1 X_{A\mf1}= i w (-{ \cos\theta} D_2 X_{A\mf1}  + \sin\theta [i \phi, X_{A\mf1}]), 
\end{split}\end{equation}
where $A=\mf1, \mf2, \mf3, \mf4$. 
These are the most general $1/8$ BPS conditions of $\cN=8$ SYM on $\mbf R\times \mbf S^2$
with the fermionic fields trivial, which preserves $\epsilon_\mf1$ with $\mf1=1,2,3,4$. 

Let us show a general solution of these $1/8$ BPS equations 
when the fields take values in Cartan subalgebra of a gauge group, 
in which case these BPS equations reduce to
\beal{
&f^{02} = - w  \cos\theta \partial^{2}\phi, \quad  
\partial_{t}\phi =-w\mu \partial_{\phi}\phi, \nn
& f_{12}= \mu\phi  -w  \sin\theta f^{01}, \quad  
\partial^{1}\phi = - w  \cos\theta f^{01},
\label{bpsf3d}\\
&(\partial_0 + {i\mu\over 2} + \mu w \partial_\phi) X_{A\mf1}=0, \quad
(\partial_1 + i w { \cos\theta} \partial_2) X_{A\mf1}=0.
\label{1/8BPSX3d}
}
The gauge field strength has to be determined by combining with the equation of motion \eqref{eomf}, 
which is now given by $\partial_\mu f^{\mu\nu} +\mu\gamma^{0\mu\nu}\partial_\mu\phi=0$.%
\footnote{
Different from the four dimensional case, the Bianchi identity becomes redundant in the three dimension.
}
Under the BPS condition \eqref{bpsf3d} this becomes 
\be
\partial_{t} f^{01}= - w\mu \partial_{\phi}  f^{01}, \quad 
\mu\partial_\phi^2\phi = \tan\theta \partial_{\theta} (\sin\theta f^{01}).
\label{eombps3d}
\ee
By carrying out dimensional reduction for \eqref{1/16BPS-Xsol}, \eqref{1/16BPS-F01sol}, 
\eqref{1/16BPS-F23sol}, we find
\beal{
X_{A\mf1} =& \sum_{s\geq0} x_{A\mf1}^{s} e^{-i \mu (s+\half) t } Y_{s, w s}(\theta,\phi), 
\label{1/8BPSXsol}\\
f^{01}
= & \sum_{s\geq\half} a_{s} c_{s} \sin ^{s-1}\theta \cos{( s(w \phi- \mu t)+ \alpha_{s})}, 
\label{1/8BPSf01}\\
\hspace{-.5cm}
\partial^{2}\phi= & \sum_{s\geq\half} a_{s} c_{s} \sin ^{s-1}\theta \sin( s(w \phi- \mu t)+ \alpha_{s}), 
}
where $x_{A\mf1}^{s}, a_s, \alpha_s$ are integral constants with the range $x_{A\mf1}^{s} \in \mbf C, a_s\geq0, 0\leq \alpha_s <2\pi$ for all $A=\mf2, \mf3, \mf4; s\geq\half$, and $Y_{s,ws}(\theta,\phi), c_s$ are given in Appendix \ref{s2}. 
Thus the real adjoint scalar field $\phi$ is determined as  
\beal{
\phi= &\phi_0-{w\over\mu} \sum_{s\geq\half} {a_{s}  \over s} c_{s} \sin ^{s}\theta \cos( s(w \phi- \mu t)+ \alpha_{s}),
\label{1/8BPSphi}
}
where $\phi_0$ is another integral constant taking real values, 
which parametrizes the vacua of the theory.%
\footnote{
If the flux quantization condition is taken into account, not only $x_{A\mf1}^{s}, a_s$ but also $\phi_0$ are quantized in a suitable way. 
} 
The other components of field strength can be obtained from \eqref{bpsf3d}. 
This is a set of general $1/8$ BPS solutions of $\cN=8$ SYM on $\mbf R\times \mbf S^2$.%
\footnote{
If we include the non-normalizable mode, the general solution \eqref{1/8BPSf01}, \eqref{1/8BPSphi} becomes 
\beal{
f^{01}
= &{a_0' \over \sin\theta} + \sum_{s\geq\half} a_{s} c_{s} \sin ^{s-1}\theta \cos{( s(w \phi - \mu t) +\alpha_{s})}, 
\label{1/8BPSf01zero}\\
\hspace{-.5cm}
\phi= &\phi_0 - {w \over \mu}\( {a_0' } \log\sin\theta +  \sum_{s\geq\half} {a_{s} \over s} c_{s} \sin ^{s}\theta \cos( s( w \phi - \mu t) + \alpha_{s})\). 
\label{1/8BPSphizero}
}
}
Note that this result is consistent with the BPS solutions obtained in \cite{Ezhuthachan:2011kf},
where BPS solutions of the $\cN=8$ SYM were obtained from those of $\cN=6$ Chern-Simons (or ABJM) theory defined on $\mbf R\times \mbf S^2$ by performing a particular scaling limit from a half BPS solution of ABJM.

We present the values of conserved charges under the BPS solution \eqref{1/8BPSXsol},  
\eqref{1/8BPSf01}, \eqref{1/8BPSphi}. 
\beal{
H
=&{1 \over g^2_3} \( 2 \mu^2\sum_{s\geq0}  (2s+1)^2  \Tr|x_{A\mf1}^{s}|^2+  \sum_{s\geq\half}  \frac{ (1+2 s)}{2 s} \Tr  (a_{s})^2\), \\
J_\theta=&0, \\
J_{\phi} =& {w \over g^2_3} \(\sum_{s\geq0}  2\mu (2s +1 ) s \Tr |x_{A\mf1}^{s}|^2+\mu^{-1} \sum_{s\geq\half} \frac{(1+2 s)}{2 s} \Tr  (a_{s})^2\),  \\
R^{\mf1}{}_{\mf1} =& {\mu \over g^2_3}  \sum_{s\geq0} (2s +1 ) \Tr|x_{A\mf1}^{s} |^2, 
}
which satisfies the BPS relation of charges given by \eqref{bpscharge3d}.  

The number of supersymmetries preserved by the $1/8$ BPS solution at each point of moduli space is given in Table \ref{table3dbps}. 
In particular this theory has degenerate vacua, which are parametrized by vacuum expectation values of the real scalar field 
denoted by $\phi_0$.%
\footnote{
The degeneracy of vacua in $\cN=8$ SYM on $\mbf R\times \mbf S^2$ can be encoded in non-trivial holonomy in the Hopf fiber direction in $\cN=4$ SYM on $\mbf R\times \mbf S^3/\mbf Z_k$,  
where the Hopf fiber direction is orbifolded \cite{Lin:2004nb,Lin:2005nh}. 
}
\begin{table}[htbp]
\begin{center}
\begin{tabular}{c|c|c}
\hline
Parameter region. &  Preserved Killing spinors. & Number of SUSY.  \\ \hline \hline
$x_{\mf1\mf2}^{s}, x_{\mf1\mf3}^{s}, x_{\mf1\mf4}^{s},a_{s}, \phi_0$: generic.   &$\eta_{\mf1}$   &  $2$  \\
   &with  $\gamma_0 \eta_{\mf1} = i w\eta_{\mf1}.$ & (${1 \over 8}$ BPS) \\
\hline
$x_{\mf1\mf3}^{s}=x_{\mf1\mf4}^{s}=0$  for $\forall s$.  & $\eta_{\mf1},\eta_{\mf2}$   &  $4$  \\
 &with  $\gamma_0 \eta_{A} = i w\eta_{A}.$ & (${1 \over 4}$ BPS) \\
\hline
 $x_{\mf1\mf2}^{s}, x_{\mf1\mf3}^{s}, x_{\mf1\mf4}^{s} =0$ for $\forall s>0$, & $\eta_{\mf1}.$   &  $4$  \\
  $a_{s}=0$  for $\forall s, r_3$.    &  & (${1 \over 4}$ BPS) \\
\hline
$x_{\mf1\mf2}^{s}=x_{\mf1\mf3}^{s}=x_{\mf1\mf4}^{s}=0$ for $\forall s$. & $\eta_{\mf1},\eta_{\mf2},\eta_{\mf3},\eta_{\mf4}$   &  $8$  \\
 &with  $\gamma_0 \eta_{A} = i w\eta_{A}$. & (${1 \over 2}$ BPS) \\
\hline
$x_{\mf1\mf2}^{s}=0$ for $\forall s>0$, &$\eta_{\mf1},\eta_{\mf2}$.  &  $8$  \\
 $x_{\mf1\mf3}^{s}=x_{\mf1\mf4}^{s}=a_{s}=0$  for $\forall s$.  &  & (${1 \over 2}$ BPS) \\
\hline
$\phi_0$: generic, & $\eta_{\mf1},\eta_{\mf2},\eta_{\mf3},\eta_{\mf4}$.   &  $16$  \\
$x_{\mf1\mf2}^{s}=x_{\mf1\mf3}^{s}=x_{\mf1\mf4}^{s}=a_s=0$ for $\forall s$. & & (Vacua) \\
\hline
\end{tabular}
\caption{ We list the number of supersymmetry of BPS solutions \eqref{1/8BPSXsol}, \eqref{1/8BPSf01}, \eqref{1/8BPSphi} which preserve $\eta_{\mf1}$ with $\gamma_0 \eta_{\mf1} = i w\eta_{\mf1}$ at each point of moduli space.
}
\label{table3dbps}
\end{center}
\end{table}

\section{Discussion} 
\label{discussion}

We have done a systematic analysis of supersymmetric states in $\cN=4$ SYM on $\mbf R\times \mbf S^3$ by setting the gaugino fields to zero. 
As a result we have found two sets of $1/16$ BPS conditions 
and we have solved them completely when the bosonic fields are valued in  
Cartan subalgebra of a gauge group. 
We have precisely counted the number of supersymmetries preserved by the BPS solutions 
varying the parameters of the solution. 
We have also obtained the most general $1/8$ BPS solution of $\cN=8$ SYM on $\mbf R\times \mbf S^2$ 
with the precise number of supersymmetries under the same assumptions by performing dimensional reduction. 

In this paper we have solved the $1/16$ BPS solutions 
assuming that they are valued in Cartan subalgebra of a gauge group. 
We pointed out that in the Cartan part 
the gauge field and the matter ones decouple 
and they can be independently excited preserving supersymmetry. 
It would be interesting to study more general BPS solutions by relaxing this assumption.
In this case one has to solve non-linear differential equations given by \eqref{bps+general} or \eqref{bps-general},
which are much more complicated and technically much harder to solve. 
Therefore it will become important to reduce the problem to a simpler one by restricting attention to a special subsector as performed in \cite{Grant:2008sk}. 

An important problem is to clarify a relation between
$1/16$ BPS states in the $\cN=4$ SYM 
and $1/16$ BPS objects in type IIB supergravity on $\mbf{AdS}_5 \times \mbf S^5$. 
It would be interesting to find out the counterparts of the BPS states constructed 
in this paper, especially those in which only electromagnetic field is turned on.  
Since charges of the supersymmetric states found in this paper can be of order $N$, 
the corresponding objects in the dual geometry can be supersymmetric (dual) giant gravitons.   
General $1/16$ BPS (dual) giant gravitons have been constructed in \cite{Kim:2006he} 
by using the same technique to construct a general $1/8$ BPS giant graviton \cite{Mikhailov:2000ya}, 
where the configuration of $1/8$ BPS giant gravitons of energy $E$ is realized by the intersection of $\mbf S^5$ and the zero locus of a polynomial of the form 
\be
\sum_{n_1+n_2+ n_3=E/R} c_{n_1n_2n_3} e^{-i E t} z_1^{n_1} z_2^{n_2} z_3^{n_3}  
\ee
where $R$ is the radius of $\mbf{AdS}_5$, $z_1, z_2, z_3$ are coordinates of $\mbf C^3$ into which $\mbf S^5$ is embedded.
(See also \cite{Mandal:2005wv,Kim:2005mw,Dhar:2005su,Mandal:2006tk,Basu:2006id,Sinha:2007ni,AliAkbari:2007jr,Ashok:2010jv}.)
Supersymmetric black holes have also been found in \cite{Gutowski:2004ez,Gutowski:2004yv,Kunduri:2006ek}, 
which have turned out to be $1/16$ BPS. 
Reproducing behaviors of these objects from the $\cN=4$ SYM side is an important issue in AdS$_5$/CFT$_4$ duality.

Another interesting direction is to study the $\cN=4$ SYM on $\mbf R\times \mbf S^3/\mbf Z_k$
by orbifolding the Hopf fiber direction 
so that the $su(2)_R$ algebra is broken and the global symmetry algebra of the system becomes $psu(2|4)$. 
Study of a class of the theories with $psu(2|4)$ symmetry is interesting 
because a family of BPS solutions with $psu(2|4)$ symmetry (called bubbling geometries) and BPS objects in ten or eleven dimensional supergravity theories has extensively studied in \cite{Lin:2004nb,Lin:2005nh,Lunin:2007ab,Lunin:2008tf}.
It would be attracting to pursue the correspondence of BPS spectra of both sides more. 
(See \cite{Okuyama:2002zn,Beisert:2002ff,Kim:2003rza,Ishiki:2006rt,Ishiki:2006yr} for the study of this direction.)

It would also be fascinating to study BPS states in other supersymmetric gauge theories 
defined on $\mbf R\times \mbf S^n$ as done in this paper, for example  
SYM on $\mbf R\times \mbf S^4$ constructed recently in \cite{Kim:2014kta}. 
(The case of $\cN=6$ Chern-Simons (ABJM) theory on $\mbf R\times \mbf S^2$ was done in \cite{Ezhuthachan:2011kf}.)

We leave these issues to future works. 

\section*{Acknowledgments} 

The author would like to thank S. Kim and S. Minwalla for useful comments 
for the first version of this paper.

\appendix 
\section{Convention}
\label{convension} 
In this appendix we collect our convention used in this paper. 
In the flat space-time, the metric is given by $g_{\mu\nu}= \text{diag}(-1,1,1,1)$
and $SO(1,3)$ gamma matrices are realized by
\be
\gamma_\mu = 
\begin{pmatrix}
 0 & \rho_\mu \\
 \rho_\mu & 0
\end{pmatrix}, \quad 
\gamma_3= 
\begin{pmatrix}
 0 &i \\
 -i & 0
\end{pmatrix}.
\label{gammarealization}
\ee
Here $\rho_0=i\sigma_2, \rho_1=\sigma_1, \rho_2=\sigma_3$, 
where $\sigma_i$ is the Pauli matrices satisfying $\sigma_i \sigma_j = \delta_{ij} +i\varepsilon_{ijk}\sigma_k$ so that $\{\gamma_\mu,\gamma_\nu\}=2g_{\mu\nu}$.
Note that $\rho_\mu$ becomes $SO(1,2)$ gamma matrices. 

In this paper we alway suppress spinor indices for simplicity.  
In a fermionic bilinear, spinor indices are contracted in the southwest-northeast manner. 
For example,  
\be
\psi^\dagger\gamma_\mu\chi= \psi^\dagger{}_{\dot\alpha}\gamma_\mu{}^{\dot\alpha}{}_\beta\chi^\beta
=C_{\dot\alpha\dot\gamma} \psi^\dagger{}^{\dot\gamma}\gamma_\mu{}^{\dot\alpha}{}_\beta\chi^\beta
\ee
where $C_{\dot\alpha\dot\gamma}$ is the charge conjugation matrix. 
Note that $\gamma_3{}^{\dot\alpha}{}_\beta= -i \delta^\alpha{}_\beta$. 
We also use notation such that 
\be
\gamma^{\mu\nu} = \half (\gamma^\mu\gamma^\nu-\gamma^\nu\gamma^\mu).
\ee

\section{Basics on $\mbf S^3 $} 
\label{s3} 

In this appendix we collect basics on $\mbf S^3$ used in this paper. 
It is convenient for us to realize $\mbf S^3$ by $SU(2)$ group. 
Any $SU(2)$ element denoted by $g$ can be parametrized by 
using the polar coordinates of $\mbf S^3$ as 
\be
g=e^{-i{\phi\over2}\sigma_3}e^{-i{\theta\over2}\sigma_2}e^{-i{\psi\over2}\sigma_3}
=\left(
\begin{array}{cc}
 \cos\frac{\theta }{2} e^{\frac{-i (\phi +\psi )}{2}} & -\sin\frac{\theta }{2} e^{\frac{i (-\phi +\psi )}{2}} \\
 \sin\frac{\theta }{2} e^{\frac{-i (-\phi +\psi )}{2}} & \cos\frac{\theta }{2} e^{\frac{i (\phi +\psi )}{2}}
\end{array}
\right)
\ee
where the parameter region is $0\leq\theta<\pi, 0\leq\phi<2\pi, 0\leq\psi<4\pi$. 
On the other hand, any point on $\mbf S^3$ with radius $r$ can embedded into $\mbf C^2$ 
by using $SU(2)$ elements  
\bea
z^i =r g^{i1}
=r \begin{pmatrix} \cos\frac{\theta }{2} e^{\frac{-i (\phi +\psi )}{2}}  \\ \sin\frac{\theta }{2} e^{\frac{-i (-\phi +\psi )}{2}} 
\end{pmatrix}
\label{s3intoc2}
\eea
where $z^1, z^2$ are complex coordinates of $\mbf C^2$. 
Therefore the metric of $\mbf S^3$ with radius $r$ is given by 
\bea
ds^2_{\mbf S^3} =|dz^1 |^2+|dz^2 |^2= {r^2 \over 4} ( d\theta^2+\sin^2\theta d\phi^2+(d\psi+ \cos\theta d\phi)^2) 
\eea
where we used \eqref{s3intoc2} to obtain the second equation. 
The metric of $\mbf R\times \mbf S^3$ is 
\be
ds^2_{\mbf R\times \mbf S^3} = - dt^2 + ds^2_{\mbf S^3}.
\ee
Therefore a local orthonormal frame is given by
\beal{
e^0 = dt, \quad 
e^{1}={r\over 2} d \theta, \quad
e^{2}={r\over 2}\sin\theta d\phi, \quad
e^{3}={r\over 2} (d \psi + \cos\theta d\phi).
}
Vielbein of this geometry reads
\bea
e^0_t=1, \quad 
e^1_{\theta}={r\over 2}, \quad
e^2_{\phi}={{r\over 2}\sin\theta }, \quad
e^3_{ \phi} ={{r\over 2} \cos\theta} ,  \quad
e^3_{ \psi} ={r\over 2}.
\eea
The inverse of vielbein is 
\bea
e^t_0=1, \quad 
e_{1}^\theta={2\over r}, \quad
e_{2}^\phi={2 \over  r \sin\theta}, \quad
e_{2}^\psi={2\over r}(- \cot\theta),  \quad
e_{3}^\psi={2\over r}.
\eea
We interchange local Lorentz indices and global space-time ones by using these. 

The 1-form connection of this geometry can be determined as follows.
\be
\omega_{21}={2\over r}(\cot\theta \, e_{2}- \half e_{3}), \quad
\omega_{13}={1 \over r} e_{2}, \quad
\omega_{23}= - {1 \over r} e_{1}.
\ee
Others components are trivial. 
This reads
\be
\omega_{2,21}={2\over r} \cot\theta, \quad
\omega_{3,21}=-{1\over r }, \quad
\omega_{2,13}={1 \over r}, \quad
\omega_{1,23}= - {1 \over r}.
\ee
The partial derivatives with the local Lorentz indices and those with the global coordinates 
are related by
\beal{
&\partial_{1} ={2\over r}\partial_{\theta}, \quad 
\partial_{2} ={2\over r}( {1\over \sin\theta} \partial_{\phi} -\cot\theta\partial_\psi) , \quad 
\partial_{3} ={2\over r} \partial_{\psi}, \\
&\partial_{\theta} ={r\over 2}\partial_{\theta}, \quad 
\partial_{\phi} ={r\over 2} (\sin\theta \partial_{2} +\cos\theta\partial_{3} ) , \quad 
\partial_{\psi} ={r\over 2} \partial_{3}.
}

A benefit to realize $\mbf S^3$ as $SU(2)$ group is that
$so(4)$ Killing action on $\mbf S^3$ is realized by the 
$su(2)_L \times su(2)_R$ algebraic action. 
\beal{
 \hat L_a g  =& -\half \sigma_a g, \quad  \hat R_a g = \half g \sigma_a. 
 }
Thanks to this definition we can easily show that 
\beal{
\left[\hat L_a,\hat L_b\right] =& i\epsilon_{abc} \hat L_c, \quad [\hat R_a, \hat R_b] = i\epsilon_{abc} \hat R_c, \quad 
[\hat L_a, \hat R_b]=0. 
}
The explicit coordinate expressions of these operators are given by 
\beal{
\hat L_1
=&i(   \sin \phi \partial_\theta-  \csc \theta \cos \phi\partial_\psi +   \cot \theta \cos \phi\partial_\phi),\nn
\hat L_2 =&i(-  \cos \phi\partial_\theta -  \csc \theta \sin \phi \partial_\psi +  \cot \theta \sin \phi \partial_\phi),\nn
\hat L_3=&- i \partial_\phi, \\
\hat R_1
=&i( \sin \psi \partial_\theta + \cot \theta \cos \psi \partial_\psi -  \csc \theta \cos \psi\partial_\phi ), \nn
 \hat R_2 =&i(  \cos \psi \partial_\theta-  \cot \theta \sin \psi  \partial_\psi +\csc \theta \sin \psi\partial_\phi ),\nn
 \hat R_3=&i  \partial_\psi.  
}
For later convenience to construct the spherical harmonics, 
we define 
\beal{ 
\hat L_\pm =& \hat L_1 \pm i\hat L_2, \quad 
\hat R_\pm = \hat R_1 \pm i\hat R_2. \quad 
\label{LRpm}
}
Then we can show that 
\beal{
\left[\hat L_3,\hat L_\pm\right] &= \pm \hat L_\pm, \quad [\hat L_+, \hat L_-]=2 \hat L_3, \\
\left[\hat R_3,\hat R_\pm\right] &= \pm \hat R_\pm, \quad [\hat R_+, \hat R_-]=2 \hat R_3, 
}
and 
\beal{
\hat L_\pm =& e^{\pm i \phi} (i\cot\theta\partial_\phi \pm \partial_\theta -i{1 \over \sin\theta} \partial_\psi), \\
\hat R_\pm =&  e^{\mp i \psi} ( i\cot\theta\partial_\psi \mp \partial_\theta - i{1 \over \sin\theta} \partial_\phi).
}

Note that 
 \beal{
 \hat L^2 = \hat R^2 =&\partial_\theta^2 + \cot\theta\partial_\theta + {1\over \sin^2\theta} (\partial_\phi^2 +\partial_\psi^2 - 2 \cos\theta\partial_\phi \partial_\psi) 
 }
 which is the same as the Laplacian of $\mbf S^3$ with radius $r=2$ acting on a scalar field. 

\subsection{The scalar spherical harmonics on $\mbf S^3$}
\label{harmonics} 

In this subsection we construct the scalar spherical harmonics on $\mbf S^3$. 
Since the spherical harmonics are roots of the $su(2)_L\times su(2)_R$ algebra, 
we can construct them by the standard algebraic method. 
\beal{
&\hat L^2 Y_{s, l_3, r_3} = \hat R^2 Y_{s, l_3, r_3} = s(s+1) Y_{s, l_3, r_3}, \quad \\
&\hat L_3 Y_{s, l_3, r_3} = l_3 Y_{s, l_3, r_3}, \quad 
\hat R_3 Y_{s, l_3, r_3} = r_3 Y_{s, l_3, r_3},
\label{eigenstates}
}
where $s$ is non-negative half integer and $|l_3|, |r_3| \leq s$. 
Firstly we determine the spherical harmonics of the highest or lowest weight of $su(2)_R$ by solving 
$R_{\pm } Y_{s, l_3, \pm s} =0 $ as well, which is given in the polar coordinates by 
\beal{
\( \partial_\theta \mp {1\over\sin\theta} l_3 - s \cot\theta  \) y_{s, l_3, \pm s}(\theta)=0
}
where $Y_{s, l_3, \pm s} = y_{s, l_3, \pm s}(\theta) e^{i(l_3\phi \mp s\psi)}$. 
The solution is given by 
\be
Y_{s, l_3, \pm s} = c_{s,l_3} \tan^{\pm l_3}\left(\frac{\theta }{2}\right) \sin ^{s}\theta  e^{i(l_3\phi \mp s\psi)},  
\label{su2Rhighest}
\ee
where $c_{s,l_3}$ is a normalization constant given by
\beal{
c_{s,l_3} ={1 \over   \pi} \sqrt {  \frac{\Gamma(2s+2)}{ 2^{2  s+1} \Gamma ( l_3+ s+1) \Gamma (- l_3+ s+1)} }
\label{csl3}
}
which we determine by demanding $ \int d^3\Omega |Y_{s, l_3, \pm s}|^2 =1$. 

In the same way, we can construct the spherical harmonics of the highest or lowest weight of $su(2)_L$ by solving 
$L_{\pm } Y_{s, \pm s,r_3} =0 $, which is 
\beal{
\( \partial_\theta \mp {1\over\sin\theta} r_3 - s \cot\theta  \) y_{s, \pm s,r_3}(\theta)=0
}
where $Y_{s, \pm s,r_3} = y_{s, \pm s,r_3}(\theta) e^{i(-r_3\psi \pm s\phi)}$. 
The solution thereof is given by 
\be
Y_{s, \pm s,r_3} = c_{s,r_3} \tan^{\pm r_3}\left(\frac{\theta }{2}\right) \sin ^{s}\theta e^{i(-r_3\psi \pm s\phi)},  
\label{su2Lhighest}
\ee
where $c_{s,r_3}$ is determined to satisfy $ \int d^3\Omega |Y_{s, \pm s,r_3} |^2 =1$ so that
\beal{
c_{s,r_3}={1 \over   \pi}  \sqrt {  \frac{\Gamma(2s+2)}{2^{2s+1} \Gamma ( r_3+ s+1) \Gamma (- r_3+ s+1)}. \label{csr3}
}
}

A general scalar spherical harmonics, $Y_{s,l_3,r_3}$, can be obtained by acting lowering operators suitably. 
\be
Y_{s,l_3,r_3} =\(\prod_{n=l_3+1}^{s} {\hat L_- \over \sqrt{s(s+1) - n(n-1)}}\) Y_{s,s,r_3}
 \ee
where we normalized the spherical harmonics to satisfy the orthonormal relation 
\be
\int d^3 \Omega (Y_{s',l_3',r_3'})^* Y_{s,l_3,r_3} = \delta_{ss'}\delta_{l_3l_3'}\delta_{r_3r_3'}.
\ee

\subsection{Reduction to $\mbf S^2$} 
\label{s2}

In this subsection we extract information on $\mbf S^2$ from the results obtained 
in the previous subsections by degenerating the Hopf fiber direction $\psi$ of $\mbf S^3$. 
The metric of $\mbf R\times \mbf S^2$ is 
\be
ds^2_{\mbf R\times \mbf S^2} =-dt^2 + {1\over \mu^2} ( d\theta^2+\sin^2\theta d\phi^2)
\ee
where $\mu$ is the inverse radius of $\mbf S^2$, which is related to the $\mbf S^3$ 
radius by $\mu={2\over r}$. 
A local orthonormal frame is
\bea
e^0 = dt, \quad
e^1= \mu^{-1}d\theta,\quad 
e^2=\mu^{-1}\sin \theta d\phi.
\eea
From this we find vielbein of $\mbf R\times \mbf S^2$ as
\beal{
&e^0_t=1, \quad 
e^1_{\theta}=\mu^{-1}, \quad
e^2_{\phi}={\mu^{-1}\sin\theta }, \nn
&e^t_0=1, \quad 
e_{1}^\theta={\mu}, \quad
e_{2}^\phi={\mu \over \sin\theta}.
\label{vielbeins2}
}
One-form connection is 
\bea
\omega^{\mbf S^2}_{21}=\mu\cot\theta e^{2}. 
\eea
Other components are zero. 

$so(3)$ Killing actions on $\mbf S^2$ are given by $\hat L_i$ with $\partial_\psi$ eliminated. 
It is automatic that $\hat L_i$ restricted on $\mbf S^2$ form $su(2)$ algebra. 
By using the operators $\hat L_i$ the $\mbf S^2$ spherical harmonics are defined as root vectors of the $su(2)$ algebra. 
\beal{
&\hat L^2 Y_{s, s_3} = s(s+1) Y_{s, s_3}, \quad \hat L_3 Y_{s, s_3} = l_3 Y_{s, s_3}, \quad 
\label{eigenstatess2}
}
where $s$ is non-negative half integer and $|s_3| \leq s$. 
The $\mbf S^2$ spherical harmonics can be obtained from $\mbf S^3$ spherical harmonics 
by taking a zero mode of $\psi$ direction. 
\be
 Y_{s, s_3}  = \sqrt{\pi \over 2} Y_{s, s_3, r_3=0}. 
\label{s2harmonicss3}
\ee
For instance, the $\mbf S^2$ spherical harmonics of the highest or lowest weight are given by
\be
Y_{s, \pm s} = c_{s} \sin ^{s}\theta  e^{\pm i s\phi},  
\label{su2Lhighests2}
\ee
where
$
c_{s}= \sqrt{\pi \over 2} c_{s,0}=  \sqrt {  \frac{\Gamma(2s+2)}{\pi 2^{2s+2} \Gamma (s+1)^2} }.
$
A general scalar spherical harmonics is given by 
\be
Y_{s, s_3} =\(\prod_{n=s_3+1}^{s} {\hat L_- \over \sqrt{s(s+1) - n(n-1)}}\) Y_{s,s}.
 \ee
These $\mbf S^2$ spherical harmonics satisfy the orthonormal relation 
\be
\int d^2 \Omega (Y_{s',s_3'})^* Y_{s, s_3} = \delta_{ss'}\delta_{s_3s_3'}
\ee
where $d^2\Omega = d\theta \sin\theta d\phi$. 

\bibliographystyle{utphys}
\bibliography{msymv2}

\end{document}